\documentclass[12pt]{article}
\usepackage{a4,epsfig,amssymb,multirow,amsmath,array}
\newcommand{\roots}{\ifmmode \sqrt{s} \else $\sqrt{s}$ \fi}
\newcommand{\tanb} {\ifmmode {\tan\beta} \else ${\tan\beta}$ \fi}

\newcommand{\evcc} {\mbox{\,eV$/c^2$}}
\newcommand{\kevcc}{\mbox{\,keV$/c^2$}}
\newcommand{\gevcc}{\mbox{\,GeV$/c^2$}}
\newcommand{\tevcc}{\mbox{\,TeV$/c^2$}}
\newcommand{\invpb}{\mbox{\,pb$^{-1}$}}

\newcommand{\neu}  {\ensuremath{\mathrm{\chi}}}

\newcommand{\grav} {\ensuremath{\mathrm{\tilde{G}}}}
\newcommand{\slep} {\ensuremath{\mathrm{\tilde{\ell}}}}
\newcommand{\stau} {\ensuremath{\mathrm{\tilde{\tau}}}}
\newcommand{\sel}  {\ensuremath{\mathrm{\tilde{e}}}}
\newcommand{\smu}  {\ensuremath{\mathrm{\tilde{\mu}}}}

\newcommand{\ee}   {\mbox{e$^{+}$e$^{-}$}}
\newcommand{\gt}{\ensuremath{>}}
\newcommand{\lt}{\ensuremath{<}}
\newcommand{\degs}{\ensuremath{^\circ}}
\newcommand{\ldet}{\mbox{$\ell_{\mathrm{detector}}$}}
\newcommand{\mtop}{\mbox{$m_{\mathrm{t}}$}}
\begin{document}
\pagestyle{empty}
\begin{center} 
{\large EUROPEAN ORGANIZATION FOR NUCLEAR RESEARCH (CERN)}  
\end{center}
\bigskip
\begin{flushright}
CERN-EP/2002-021\\
%{\bf Draft 3}\\
March 13, 2002 \\
%\begin{center} {\LARGE PRELIMINARY}  \end{center}
\bigskip
\end{flushright}
\vspace{2cm}

\begin{center}
 \mathversion{bold}

{\LARGE\bf Search for gauge mediated SUSY breaking topologies in $\ee$
collisions at centre--of--mass energies up to 209 GeV}
\mathversion{normal}
  \vskip 1.2cm
  {\bf The ALEPH Collaboration} $^*)$\\ 
  \vskip 1.2cm
  {\bf Abstract}
\end{center}
\medskip
%
% ABSTRACT 
%
\small
A total of 628$\invpb$ of data collected with the ALEPH detector 
at centre-of-mass energies from 189 to 209\,GeV is analysed in the
search for gauge mediated SUSY breaking (GMSB) topologies. 
These topologies include 
two acoplanar photons, non-pointing single photons,
acoplanar leptons, large impact parameter leptons, detached slepton
decay vertices, heavy stable charged sleptons and multi-leptons plus
missing energy final states.
No evidence is found for new phenomena, and lower limits on
masses of supersymmetric particles are derived. 
A scan of a minimal GMSB parameter space is performed and 
lower limits are set for the next-to-lightest supersymmetric 
particle (NLSP) mass at 54$\gevcc$ and for the mass scale parameter
$\Lambda$ at 10$\tevcc$, independently of the NLSP lifetime.
Including the results from the neutral Higgs boson searches,
a NLSP mass limit of 77$\gevcc$ is obtained and values of $\Lambda$ up
to 16$\tevcc$ are excluded.
\begin{center}
\vspace{2cm}
{\it (Submitted to European Physics Journal C)} \\
\vspace{1cm}
\end{center}
\vfill
\font\ninerm=cmr9
\noindent
---------------------------------------------\hfil\break
{\ninerm $^*)$ See next pages for the list of authors}
%
%%%%%%%%%%%%%%%%%%%%%%%%%%%%%%%%%%%%%%%%%%%%%%%%%%%%%%%%%%%%%%%%%%%%%%%%%%%%%
%% Get author list from: /afs/cern.ch/user/a/alephsec/public/authlist/authb.tex
%------------------------------------------------------------------------
% authob12pt.tex
% authors' list for papers at LEP 1.5 and 2 energies
%-----------------------------------------------------------------------
\pagestyle{empty}
\newpage
\small
%
% remember the old settings
%
\newlength{\saveparskip}
\newlength{\savetextheight}
\newlength{\savetopmargin}
\newlength{\savetextwidth}
\newlength{\saveoddsidemargin}
\newlength{\savetopsep}
\setlength{\saveparskip}{\parskip}
\setlength{\savetextheight}{\textheight}
\setlength{\savetopmargin}{\topmargin}
\setlength{\savetextwidth}{\textwidth}
\setlength{\saveoddsidemargin}{\oddsidemargin}
\setlength{\savetopsep}{\topsep}
%
% text dimensions for the author list
%
\setlength{\parskip}{0.0cm}
\setlength{\textheight}{25.0cm}
\setlength{\topmargin}{-1.5cm}
\setlength{\textwidth}{16 cm}
\setlength{\oddsidemargin}{-0.0cm}
\setlength{\topsep}{1mm}
\pretolerance=10000
%%\begin{document}
%\centerline{EUROPEAN ORGANIZATION FOR NUCLEAR RESEARCH}
%\centerline{EUROPEAN LABORATORY FOR PARTICLE PHYSICS (CERN)}
%\vspace{1cm}
%\begin{flushright}CERN-EP-2000-
%\20 February 2002 - last update
%\end{flushright}
\centerline{\large\bf The ALEPH Collaboration}
\footnotesize
\vspace{0.5cm}
{\raggedbottom
\begin{sloppypar}
\samepage\noindent
A.~Heister,
S.~Schael
\nopagebreak
\begin{center}
\parbox{15.5cm}{\sl\samepage
Physikalisches Institut das RWTH-Aachen, D-52056 Aachen, Germany}
\end{center}\end{sloppypar}
\vspace{2mm}
\begin{sloppypar}
\noindent
R.~Barate,
R.~Bruneli\`ere,
I.~De~Bonis,
D.~Decamp,
C.~Goy,
S.~Jezequel,
J.-P.~Lees,
F.~Martin,
E.~Merle,
\mbox{M.-N.~Minard},
B.~Pietrzyk,
B.~Trocm\'e
\nopagebreak
\begin{center}
\parbox{15.5cm}{\sl\samepage
Laboratoire de Physique des Particules (LAPP), IN$^{2}$P$^{3}$-CNRS,
F-74019 Annecy-le-Vieux Cedex, France}
\end{center}\end{sloppypar}
\vspace{2mm}
\begin{sloppypar}
\noindent
G.~Boix,$^{25}$
S.~Bravo,
M.P.~Casado,
M.~Chmeissani,
J.M.~Crespo,
E.~Fernandez,
M.~Fernandez-Bosman,
Ll.~Garrido,$^{15}$
E.~Graug\'{e}s,
J.~Lopez,
M.~Martinez,
G.~Merino,
R.~Miquel,$^{4}$
Ll.M.~Mir,$^{4}$
A.~Pacheco,
D.~Paneque,
H.~Ruiz
\nopagebreak
\begin{center}
\parbox{15.5cm}{\sl\samepage
Institut de F\'{i}sica d'Altes Energies, Universitat Aut\`{o}noma
de Barcelona, E-08193 Bellaterra (Barcelona), Spain$^{7}$}
\end{center}\end{sloppypar}
\vspace{2mm}
\begin{sloppypar}
\noindent
A.~Colaleo,
D.~Creanza,
N.~De~Filippis,
M.~de~Palma,
G.~Iaselli,
G.~Maggi,
M.~Maggi,
S.~Nuzzo,
A.~Ranieri,
G.~Raso,$^{24}$
F.~Ruggieri,
G.~Selvaggi,
L.~Silvestris,
P.~Tempesta,
A.~Tricomi,$^{3}$
G.~Zito
\nopagebreak
\begin{center}
\parbox{15.5cm}{\sl\samepage
Dipartimento di Fisica, INFN Sezione di Bari, I-70126 Bari, Italy}
\end{center}\end{sloppypar}
\vspace{2mm}
\begin{sloppypar}
\noindent
X.~Huang,
J.~Lin,
Q. Ouyang,
T.~Wang,
Y.~Xie,
R.~Xu,
S.~Xue,
J.~Zhang,
L.~Zhang,
W.~Zhao
\nopagebreak
\begin{center}
\parbox{15.5cm}{\sl\samepage
Institute of High Energy Physics, Academia Sinica, Beijing, The People's
Republic of China$^{8}$}
\end{center}\end{sloppypar}
\vspace{2mm}
\begin{sloppypar}
\noindent
D.~Abbaneo,
P.~Azzurri,
T.~Barklow,$^{30}$
O.~Buchm\"uller,$^{30}$
M.~Cattaneo,
F.~Cerutti,
B.~Clerbaux,
H.~Drevermann,
R.W.~Forty,
M.~Frank,
F.~Gianotti,
T.C.~Greening,$^{26}$
J.B.~Hansen,
J.~Harvey,
D.E.~Hutchcroft,
P.~Janot,
B.~Jost,
M.~Kado,$^{2}$
P.~Mato,
A.~Moutoussi,
F.~Ranjard,
L.~Rolandi,
D.~Schlatter,
G.~Sguazzoni,
W.~Tejessy,
F.~Teubert,
A.~Valassi,
I.~Videau,
J.J.~Ward
\nopagebreak
\begin{center}
\parbox{15.5cm}{\sl\samepage
European Laboratory for Particle Physics (CERN), CH-1211 Geneva 23,
Switzerland}
\end{center}\end{sloppypar}
\vspace{2mm}
\begin{sloppypar}
\noindent
F.~Badaud,
S.~Dessagne,
A.~Falvard,$^{20}$
D.~Fayolle,
P.~Gay,
J.~Jousset,
B.~Michel,
S.~Monteil,
D.~Pallin,
J.M.~Pascolo,
P.~Perret
\nopagebreak
\begin{center}
\parbox{15.5cm}{\sl\samepage
Laboratoire de Physique Corpusculaire, Universit\'e Blaise Pascal,
IN$^{2}$P$^{3}$-CNRS, Clermont-Ferrand, F-63177 Aubi\`{e}re, France}
\end{center}\end{sloppypar}
\vspace{2mm}
\begin{sloppypar}
\noindent
J.D.~Hansen,
J.R.~Hansen,
P.H.~Hansen,
B.S.~Nilsson
\nopagebreak
\begin{center}
\parbox{15.5cm}{\sl\samepage
Niels Bohr Institute, 2100 Copenhagen, DK-Denmark$^{9}$}
\end{center}\end{sloppypar}
\vspace{2mm}
\begin{sloppypar}
\noindent
A.~Kyriakis,
C.~Markou,
E.~Simopoulou,
A.~Vayaki,
K.~Zachariadou
\nopagebreak
\begin{center}
\parbox{15.5cm}{\sl\samepage
Nuclear Research Center Demokritos (NRCD), GR-15310 Attiki, Greece}
\end{center}\end{sloppypar}
\vspace{2mm}
\begin{sloppypar}
\noindent
A.~Blondel,$^{12}$
\mbox{J.-C.~Brient},
F.~Machefert,
A.~Roug\'{e},
M.~Swynghedauw,
R.~Tanaka
\linebreak
H.~Videau
\nopagebreak
\begin{center}
\parbox{15.5cm}{\sl\samepage
Laboratoire de Physique Nucl\'eaire et des Hautes Energies, Ecole
Polytechnique, IN$^{2}$P$^{3}$-CNRS, \mbox{F-91128} Palaiseau Cedex, France}
\end{center}\end{sloppypar}
\vspace{2mm}
\begin{sloppypar}
\noindent
V.~Ciulli,
E.~Focardi,
G.~Parrini
\nopagebreak
\begin{center}
\parbox{15.5cm}{\sl\samepage
Dipartimento di Fisica, Universit\`a di Firenze, INFN Sezione di Firenze,
I-50125 Firenze, Italy}
\end{center}\end{sloppypar}
\vspace{2mm}
\begin{sloppypar}
\noindent
A.~Antonelli,
M.~Antonelli,
G.~Bencivenni,
F.~Bossi,
P.~Campana,
G.~Capon,
V.~Chiarella,
P.~Laurelli,
G.~Mannocchi,$^{5}$
F.~Murtas,
G.P.~Murtas,
L.~Passalacqua
\nopagebreak
\begin{center}
\parbox{15.5cm}{\sl\samepage
Laboratori Nazionali dell'INFN (LNF-INFN), I-00044 Frascati, Italy}
\end{center}\end{sloppypar}
\vspace{2mm}
%\pagebreak
\begin{sloppypar}
\noindent
A.~Halley,
J.~Kennedy,
J.G.~Lynch,
P.~Negus,
V.~O'Shea,
A.S.~Thompson
\nopagebreak
\begin{center}
\parbox{15.5cm}{\sl\samepage
Department of Physics and Astronomy, University of Glasgow, Glasgow G12
8QQ,United Kingdom$^{10}$}
\end{center}\end{sloppypar}
\vspace{2mm}
%\pagebreak
\begin{sloppypar}
\noindent
S.~Wasserbaech
\nopagebreak
\begin{center}
\parbox{15.5cm}{\sl\samepage
Department of Physics, Haverford College, Haverford, PA 19041-1392, U.S.A.}
\end{center}\end{sloppypar}
\vspace{2mm}
%\pagebreak
\begin{sloppypar}
\noindent
R.~Cavanaugh,$^{33}$
S.~Dhamotharan,$^{34}$
C.~Geweniger,
P.~Hanke,
V.~Hepp,
E.E.~Kluge,
G.~Leibenguth,
A.~Putzer,
H.~Stenzel,
K.~Tittel,
M.~Wunsch$^{19}$
\nopagebreak
\begin{center}
\parbox{15.5cm}{\sl\samepage
Kirchhoff-Institut f\"ur Physik, Universit\"at Heidelberg, D-69120
Heidelberg, Germany$^{16}$}
\end{center}\end{sloppypar}
\vspace{2mm}
\begin{sloppypar}
\noindent
R.~Beuselinck,
W.~Cameron,
G.~Davies,
P.J.~Dornan,
M.~Girone,$^{1}$
R.D.~Hill,
N.~Marinelli,
J.~Nowell,
S.A.~Rutherford,
J.K.~Sedgbeer,
J.C.~Thompson,$^{14}$
R.~White
\nopagebreak
\begin{center}
\parbox{15.5cm}{\sl\samepage
Department of Physics, Imperial College, London SW7 2BZ,
United Kingdom$^{10}$}
\end{center}\end{sloppypar}
\vspace{2mm}
\begin{sloppypar}
\noindent
V.M.~Ghete,
P.~Girtler,
E.~Kneringer,
D.~Kuhn,
G.~Rudolph
\nopagebreak
\begin{center}
\parbox{15.5cm}{\sl\samepage
Institut f\"ur Experimentalphysik, Universit\"at Innsbruck, A-6020
Innsbruck, Austria$^{18}$}
\end{center}\end{sloppypar}
\vspace{2mm}
\begin{sloppypar}
\noindent
E.~Bouhova-Thacker,
C.K.~Bowdery,
D.P.~Clarke,
G.~Ellis,
A.J.~Finch,
F.~Foster,
G.~Hughes,
R.W.L.~Jones,
M.R.~Pearson,
N.A.~Robertson,
M.~Smizanska
\nopagebreak
\begin{center}
\parbox{15.5cm}{\sl\samepage
Department of Physics, University of Lancaster, Lancaster LA1 4YB,
United Kingdom$^{10}$}
\end{center}\end{sloppypar}
\vspace{2mm}
\begin{sloppypar}
\noindent
O.~van~der~Aa,
C.~Delaere,
V.~Lemaitre
\nopagebreak
\begin{center}
\parbox{15.5cm}{\sl\samepage
Institut de Physique Nucl\'eaire, D\'epartement de Physique, Universit\'e Catholique de Louvain, 1348 Louvain-la-Neuve, Belgium}
\end{center}\end{sloppypar}
\vspace{2mm}
\begin{sloppypar}
\noindent
U.~Blumenschein,
F.~H\"olldorfer,
K.~Jakobs,
F.~Kayser,
K.~Kleinknecht,
A.-S.~M\"uller,
G.~Quast,$^{6}$
B.~Renk,
H.-G.~Sander,
S.~Schmeling,
H.~Wachsmuth,
C.~Zeitnitz,
T.~Ziegler
\nopagebreak
\begin{center}
\parbox{15.5cm}{\sl\samepage
Institut f\"ur Physik, Universit\"at Mainz, D-55099 Mainz, Germany$^{16}$}
\end{center}\end{sloppypar}
\vspace{2mm}
\begin{sloppypar}
\noindent
A.~Bonissent,
P.~Coyle,
C.~Curtil,
A.~Ealet,
D.~Fouchez,
P.~Payre,
A.~Tilquin
\nopagebreak
\begin{center}
\parbox{15.5cm}{\sl\samepage
Centre de Physique des Particules de Marseille, Univ M\'editerran\'ee,
IN$^{2}$P$^{3}$-CNRS, F-13288 Marseille, France}
\end{center}\end{sloppypar}
\vspace{2mm}
\begin{sloppypar}
\noindent
F.~Ragusa
\nopagebreak
\begin{center}
\parbox{15.5cm}{\sl\samepage
Dipartimento di Fisica, Universit\`a di Milano e INFN Sezione di
Milano, I-20133 Milano, Italy.}
\end{center}\end{sloppypar}
\vspace{2mm}
\begin{sloppypar}
\noindent
A.~David,
H.~Dietl,
G.~Ganis,$^{27}$
K.~H\"uttmann,
G.~L\"utjens,
W.~M\"anner,
\mbox{H.-G.~Moser},
R.~Settles,
G.~Wolf
\nopagebreak
\begin{center}
\parbox{15.5cm}{\sl\samepage
Max-Planck-Institut f\"ur Physik, Werner-Heisenberg-Institut,
D-80805 M\"unchen, Germany\footnotemark[16]}
\end{center}\end{sloppypar}
\vspace{2mm}
\begin{sloppypar}
\noindent
J.~Boucrot,
O.~Callot,
M.~Davier,
L.~Duflot,
\mbox{J.-F.~Grivaz},
Ph.~Heusse,
A.~Jacholkowska,$^{32}$
C.~Loomis,
L.~Serin,
\mbox{J.-J.~Veillet},
J.-B.~de~Vivie~de~R\'egie,$^{28}$
C.~Yuan
\nopagebreak
\begin{center}
\parbox{15.5cm}{\sl\samepage
Laboratoire de l'Acc\'el\'erateur Lin\'eaire, Universit\'e de Paris-Sud,
IN$^{2}$P$^{3}$-CNRS, F-91898 Orsay Cedex, France}
\end{center}\end{sloppypar}
\vspace{2mm}
\begin{sloppypar}
\noindent
%\samepage
G.~Bagliesi,
T.~Boccali,
L.~Fo\`a,
A.~Giammanco,
A.~Giassi,
F.~Ligabue,
A.~Messineo,
F.~Palla,
G.~Sanguinetti,
A.~Sciab\`a,
R.~Tenchini,$^{1}$
A.~Venturi,$^{1}$
P.G.~Verdini
\samepage
\begin{center}
\parbox{15.5cm}{\sl\samepage
Dipartimento di Fisica dell'Universit\`a, INFN Sezione di Pisa,
e Scuola Normale Superiore, I-56010 Pisa, Italy}
\end{center}\end{sloppypar}
\vspace{2mm}
\begin{sloppypar}
\noindent
O.~Awunor,
G.A.~Blair,
G.~Cowan,
A.~Garcia-Bellido,
M.G.~Green,
L.T.~Jones,
T.~Medcalf,
A.~Misiejuk,
J.A.~Strong,
P.~Teixeira-Dias
\nopagebreak
\begin{center}
\parbox{15.5cm}{\sl\samepage
Department of Physics, Royal Holloway \& Bedford New College,
University of London, Egham, Surrey TW20 OEX, United Kingdom$^{10}$}
\end{center}\end{sloppypar}
\vspace{2mm}
\begin{sloppypar}
\noindent
R.W.~Clifft,
T.R.~Edgecock,
P.R.~Norton,
I.R.~Tomalin
\nopagebreak
\begin{center}
\parbox{15.5cm}{\sl\samepage
Particle Physics Dept., Rutherford Appleton Laboratory,
Chilton, Didcot, Oxon OX11 OQX, United Kingdom$^{10}$}
\end{center}\end{sloppypar}
\vspace{2mm}
%\pagebreak
\begin{sloppypar}
\noindent
\mbox{B.~Bloch-Devaux},
D.~Boumediene,
P.~Colas,
B.~Fabbro,
E.~Lan\c{c}on,
\mbox{M.-C.~Lemaire},
E.~Locci,
P.~Perez,
J.~Rander,
\mbox{J.-F.~Renardy},
A.~Rosowsky,
P.~Seager,$^{13}$
A.~Trabelsi,$^{21}$
B.~Tuchming,
B.~Vallage
\nopagebreak
\begin{center}
\parbox{15.5cm}{\sl\samepage
CEA, DAPNIA/Service de Physique des Particules,
CE-Saclay, F-91191 Gif-sur-Yvette Cedex, France$^{17}$}
\end{center}\end{sloppypar}
%\nopagebreak
\vspace{2mm}
\begin{sloppypar}
\noindent
N.~Konstantinidis,
A.M.~Litke,
G.~Taylor
\nopagebreak
\begin{center}
\parbox{15.5cm}{\sl\samepage
Institute for Particle Physics, University of California at
Santa Cruz, Santa Cruz, CA 95064, USA$^{22}$}
\end{center}\end{sloppypar}
%\pagebreak
\vspace{2mm}
\begin{sloppypar}
\noindent
C.N.~Booth,
S.~Cartwright,
F.~Combley,$^{31}$
P.N.~Hodgson,
M.~Lehto,
L.F.~Thompson
\nopagebreak
\begin{center}
\parbox{15.5cm}{\sl\samepage
Department of Physics, University of Sheffield, Sheffield S3 7RH,
United Kingdom$^{10}$}
\end{center}\end{sloppypar}
\vspace{2mm}
\begin{sloppypar}
\noindent
K.~Affholderbach,$^{23}$
A.~B\"ohrer,
S.~Brandt,
C.~Grupen,
J.~Hess,
A.~Ngac,
G.~Prange,
U.~Sieler
\nopagebreak
\begin{center}
\parbox{15.5cm}{\sl\samepage
Fachbereich Physik, Universit\"at Siegen, D-57068 Siegen, Germany$^{16}$}
\end{center}\end{sloppypar}
\vspace{2mm}
\begin{sloppypar}
\noindent
C.~Borean,
G.~Giannini
\nopagebreak
\begin{center}
\parbox{15.5cm}{\sl\samepage
Dipartimento di Fisica, Universit\`a di Trieste e INFN Sezione di Trieste,
I-34127 Trieste, Italy}
\end{center}\end{sloppypar}
\vspace{2mm}
\begin{sloppypar}
\noindent
H.~He,
J.~Putz,
J.~Rothberg
\nopagebreak
\begin{center}
\parbox{15.5cm}{\sl\samepage
Experimental Elementary Particle Physics, University of Washington, Seattle,
WA 98195 U.S.A.}
\end{center}\end{sloppypar}
\vspace{2mm}
\begin{sloppypar}
\noindent
S.R.~Armstrong,
K.~Berkelman,
K.~Cranmer,
D.P.S.~Ferguson,
Y.~Gao,$^{29}$
S.~Gonz\'{a}lez,
O.J.~Hayes,
H.~Hu,
S.~Jin,
J.~Kile,
P.A.~McNamara III,
J.~Nielsen,
Y.B.~Pan,
\mbox{J.H.~von~Wimmersperg-Toeller}, 
W.~Wiedenmann,
J.~Wu,
Sau~Lan~Wu,
X.~Wu,
G.~Zobernig
\nopagebreak
\begin{center}
\parbox{15.5cm}{\sl\samepage
Department of Physics, University of Wisconsin, Madison, WI 53706,
USA$^{11}$}
\end{center}\end{sloppypar}
\vspace{2mm}
\begin{sloppypar}
\noindent
G.~Dissertori
\nopagebreak
\begin{center}
\parbox{15.5cm}{\sl\samepage
Institute for Particle Physics, ETH H\"onggerberg, 8093 Z\"urich,
Switzerland.}
\end{center}\end{sloppypar}
}
\footnotetext[1]{Also at CERN, 1211 Geneva 23, Switzerland.}
\footnotetext[2]{Now at Fermilab, PO Box 500, MS 352, Batavia, IL 60510, USA}
\footnotetext[3]{Also at Dipartimento di Fisica di Catania and INFN Sezione di
 Catania, 95129 Catania, Italy.}
\footnotetext[4]{Now at LBNL, Berkeley, CA 94720, U.S.A.}
\footnotetext[5]{Also Istituto di Cosmo-Geofisica del C.N.R., Torino,
Italy.}
\footnotetext[6]{Now at Institut f\"ur Experimentelle Kernphysik, Universit\"at Karlsruhe, 76128 Karlsruhe, Germany.}
\footnotetext[7]{Supported by CICYT, Spain.}
\footnotetext[8]{Supported by the National Science Foundation of China.}
\footnotetext[9]{Supported by the Danish Natural Science Research Council.}
\footnotetext[10]{Supported by the UK Particle Physics and Astronomy Research
Council.}
\footnotetext[11]{Supported by the US Department of Energy, grant
DE-FG0295-ER40896.}
\footnotetext[12]{Now at Departement de Physique Corpusculaire, Universit\'e de
Gen\`eve, 1211 Gen\`eve 4, Switzerland.}
\footnotetext[13]{Supported by the Commission of the European Communities,
contract ERBFMBICT982874.}
\footnotetext[14]{Supported by the Leverhulme Trust.}
\footnotetext[15]{Permanent address: Universitat de Barcelona, 08208 Barcelona,
Spain.}
\footnotetext[16]{Supported by Bundesministerium f\"ur Bildung
und Forschung, Germany.}
\footnotetext[17]{Supported by the Direction des Sciences de la
Mati\`ere, C.E.A.}
\footnotetext[18]{Supported by the Austrian Ministry for Science and Transport.}
\footnotetext[19]{Now at SAP AG, 69185 Walldorf, Germany}
\footnotetext[20]{Now at Groupe d' Astroparticules de Montpellier, Universit\'e de Montpellier II, 34095 Montpellier, France.}
\footnotetext[21]{Now at D\'epartement de Physique, Facult\'e des Sciences de Tunis, 1060 Le Belv\'ed\`ere, Tunisia.}
\footnotetext[22]{Supported by the US Department of Energy,
grant DE-FG03-92ER40689.}
\footnotetext[23]{Now at Skyguide, Swissair Navigation Services, Geneva, Switzerland.}
\footnotetext[24]{Also at Dipartimento di Fisica e Tecnologie Relative, Universit\`a di Palermo, Palermo, Italy.}
\footnotetext[25]{Now at McKinsey and Compagny, Avenue Louis Casal 18, 1203 Geneva, Switzerland.}
\footnotetext[26]{Now at Honeywell, Phoenix AZ, U.S.A.}
\footnotetext[27]{Now at INFN Sezione di Roma II, Dipartimento di Fisica, Universit\`a di Roma Tor Vergata, 00133 Roma, Italy.}
\footnotetext[28]{Now at Centre de Physique des Particules de Marseille, Univ M\'editerran\'ee, F-13288 Marseille, France.}
\footnotetext[29]{Also at Department of Physics, Tsinghua University, Beijing, The People's Republic of China.}
\footnotetext[30]{Now at SLAC, Stanford, CA 94309, U.S.A.}
\footnotetext[31]{Deceased.}
\footnotetext[32]{Also at Groupe d' Astroparticules de Montpellier, Universit\'e de Montpellier II, 34095 Montpellier, France.}  
\footnotetext[33]{Now at University of Florida, Department of Physics, Gainesville, Florida 32611-8440, USA}
\footnotetext[34]{Now at BNP Paribas, 60325 Frankfurt am Mainz, Germany}
\setlength{\parskip}{\saveparskip}
\setlength{\textheight}{\savetextheight}
\setlength{\topmargin}{\savetopmargin}
\setlength{\textwidth}{\savetextwidth}
\setlength{\oddsidemargin}{\saveoddsidemargin}
\setlength{\topsep}{\savetopsep}
\normalsize
\newpage
\pagestyle{plain}
\setcounter{page}{1}

%%%%%%%%%%%%%%%%%%%%%%%%%%%%%%%%%%%%%%%%%%%%%%%%%%%%%%%%%%%%%%%%%%%%%%%%%%%%%
\newpage
\pagestyle{plain}
\setcounter{page}{1}
\normalsize
%
% INTRODUCTION
%
\section{Introduction}
\label{intro}
If supersymmetry (SUSY) were an exact symmetry, 
the new SUSY particles would be degenerate in mass with their
Standard Model (SM) partners. As no experimental evidence has been
found to prove the existence of SUSY particles, supersymmetry 
must be a broken symmetry.
In Gauge Mediated SUSY Breaking (GMSB) models, supersymmetry is broken
at a high energy scale in a hidden or ``secluded'' sector and is
then propagated down to the visible sector via the SM gauge 
interactions~\cite{theory}.
The main motivation for GMSB models lies in the fact that they can
easily cope with the experimental absence of flavour changing neutral
currents (FCNC).
Gauge interactions are flavour blind and the scale at which SUSY breaking is
mediated is expected to be well below the scale at which
flavour symmetry should be broken.

From the phenomenological point of view the main difference with
respect to gravity
mediated SUSY breaking models (SUGRA)~\cite{sugra} 
is that in GMSB the lightest supersymmetric
particle (LSP) is the gravitino (\grav) which couples very weakly to the other
particles. Assuming R-parity conservation, SUSY particles are pair produced
in $\ee$ collisions and subsequently decay to their SM partner plus gravitinos.
Another important characteristic of these models is that the next-to-lightest
supersymmetric particle (NLSP) is, in general, either the lightest
neutralino $\neu$ or the sleptons $\slep$.   
In GMSB models a non-negligible mixing between $\stau_{\mathrm{L}}$ and 
$\stau_{\mathrm{R}}$ states is expected for moderate and large values of \tanb 
(the ratio of the vacuum expectation values of the two Higgs
doublets) or large values of $|\mu|$ (the Higgs mixing mass term). If
the stau mixing is large, the lightest stau $\stau_1$ becomes lighter 
than the other sleptons, and also possibly lighter than the neutralino,
being then the only NLSP. 

The lifetime of the NLSP depends on the gravitino mass (or equivalently on
the SUSY breaking scale $\sqrt{F}$ which is proportional to it)~\cite{theory}: 
\begin{equation} \label{ctaunlsp}
c\tau_{\mathrm{NLSP}} \approx \frac{0.01}{\kappa_{\gamma}} 
\left ( \frac{100\,\mathrm{GeV}}{m_{\mathrm{NLSP}}} \right )^5 
\left ( \frac{m_{\grav}}{2.4\,\mathrm{eV}} \right )^2 \mathrm{cm}
\end{equation} 
where $\kappa_{\gamma}$ is the bino component of the $\neu$, and
$\kappa_{\gamma}=1$ for a $\slep$ NLSP.
When cosmological considerations are taken into account, an
upper limit is placed on the gravitino mass~\cite{cosmo}: 
$m_{\grav}~\lesssim~1\kevcc$ $(\sqrt{F}~\lesssim~2000$\,TeV). 
Thus the gravitino mass can range from $\mathcal{O}(10^{-2})\evcc$ to 1$\kevcc$,
which practically implies that any NLSP decay length is allowed.
For this reason topological searches able to identify long-lived
or even stable NLSP's have been developed by the ALEPH
collaboration \cite{gg97,slep97,oldcasc}.

A previous compilation of all GMSB searches carried out by ALEPH
exists with data at $\roots = 189$\,GeV~\cite{gmsbpaper}. 
In this paper the results of the GMSB searches for all
data collected at \roots up to 209\,GeV are summarised.
Other LEP and Tevatron experiments have reported their results in
Refs.~\cite{delphigmsb,opalgmsb,tevatrongmsb}.
%including up to 444$invpb$ of new data since 1998. 

The organisation of this paper is the following. 
A brief description of the ALEPH detector is presented in Section~\ref{detector}.
In Section~\ref{searches}, the experimental topologies are reviewed
and limits on sparticle masses are reported. 
An update on four-lepton final states and
the new selection for four-lepton final states when sleptons have
lifetime are described in Section~\ref{cascade}. 
In Section~\ref{scan} the scan on a
minimal set of GMSB parameters is presented. The sensitivity of these parameters
to the different search exclusions is analysed and lower limits on
the NLSP mass and the mass scale parameter $\Lambda$ are derived.
%%%%%%%%%%%%%%%%%%%%%%%%%%%%%%%%%%%%%%%%%%%%%%%%%%%%%%%%%%%%%%%%%%%%%%
% ALEPH Detector
%
\section{The ALEPH detector and data samples}
\label{detector}
A detailed description of the ALEPH detector can be found in Ref.~\cite{Alnim},
and an account of its performance as well as a description of the
standard analysis algorithms can be found in Ref.~\cite{Alperf}.
Only a brief overview is given here.

Charged particle tracks are measured by a silicon vertex detector (VDET), 
a multiwire drift chamber (ITC) and a time projection chamber (TPC).
The VDET has a length of approximately 40\,cm with two concentric layers of 
silicon wafers at average radii of 6.5 and 11.3\,cm.  
The ITC consists of eight drift chamber layers of 2\,m length between an inner 
radius of 16\,cm and an outer radius of 26\,cm. 
The TPC measures up to 21 space points in the radial range from 30\,cm to 
180\,cm and has an overall length of 4.4\,m.      
These detectors are immersed in an axial magnetic field of 1.5\,T and together
achieve a transverse momentum resolution 
$\sigma(p_{\mathrm{T}})/p_{\mathrm{T}}=0.0006 p_{\mathrm{T}} \oplus 0.005$ ($p_{\mathrm{T}}$ in GeV/$c$).
The TPC also provides up to 338 measurements of the ionisation energy loss. 
It is surrounded by the electromagnetic calorimeter (ECAL), which covers
the angular range $|{\cos\theta}|<0.98$. 
The ECAL is finely segmented in projective towers of approximately 
$0.9^{\circ}$ by $0.9^{\circ}$ which are read out in three segments of depth.
The energy resolution is $\sigma(E)/E = 0.18/\sqrt{E} + 0.009$ ($E$ in GeV).
The iron return yoke is instrumented with streamer tubes acting as a hadron
calorimeter (HCAL) and covers polar angles down to 110\,mrad.
Surrounding the HCAL are two additional double layers of streamer tubes
called muon chambers.
The luminosity monitors (LCAL and SICAL) extend the calorimetric coverage
down to polar angles of 34\,mrad. 

Using the energy flow algorithm described in Ref.~\cite{Alperf}, the 
measurements of the tracking detectors and the calorimeters are combined into 
``objects'' classified as charged particles, photons, and neutral hadrons.
A {\it good track\/} is defined as a charged particle track originating from 
the interaction region (with transverse impact parameter $|d_{0}|<$ 1\,cm and
longitudinal impact parameter $|z_{0}|<$ 5\,cm), having at least four TPC hits,
a transverse momentum greater than 200\,MeV/$c$ and a minimum polar angle of 
$18.2^\circ$.
In order to get the correct charged multiplicity, photon 
conversions are reconstructed with a pair-finding 
algorithm~\cite{Alperf}.
Electrons are identified by comparing the energy deposit in ECAL to 
the momentum measured in the tracking system, by using the shower profile in the 
electromagnetic calorimeter and by the measurement of the specific
ionisation in the TPC. 
The tagging of muons makes use of the hit patterns in HCAL and the muon
chambers.

The data samples used in this paper, collected by the ALEPH detector 
from 1998 to 2000, are given in Table~\ref{lumin}.
\begin{table}[b]
\begin{center}
\caption[]{\label{lumin} \small Average centre-of-mass energy and
corresponding luminosities of the analysed data sample.}
%for data collected by the ALEPH detector from 1998 to 2000.}
\vspace*{0.2cm}
\begin{tabular}{|c||c|c|c|c|c|c|c|} \hline
Year  & 1998 & \multicolumn{4}{c|}{1999} & \multicolumn{2}{c|}{2000} \\ \hline
$\langle \sqrt{s} \rangle$ (GeV) & 188.6 & 191.6 & 195.5 & 199.5 & 201.6 & 205.0 & 206.7 \\ \hline
$\int \mathcal{L}$dt ($\invpb$)         & 173.6 &  28.9 &  79.9 &  87.0 &  44.4 &  79.5 & 134.3 \\ \hline
\end{tabular}
\end{center}
\end{table}
All selections were developed using Monte Carlo techniques. 
Simulated samples corresponding to at least ten times the collected 
luminosity of all major background processes have been generated. 
A detailed list of the Monte Carlo generators used can be found in 
Refs.~\cite{slepba,photons}. Signal samples were simulated with {\tt
SUSYGEN}~\cite{susygen}.
The position of the most important cuts was determined using the $\bar{N}_{95}$
prescription~\cite{nbar95}, which corresponds to the minimisation of the
expected 95$\%$ confidence level upper limit on the number of signal
events, under the hypothesis that no signal is present in the data.

%%%%%%%%%%%%%%%%%%%%%%%%%%%%%%%%%%%%%%%%%%%%%%%%%%%%%%%%%%%%%%%%%%%%%%
% Experimental Topologies 
%%%%%%%%%%%%%%%%%%%%%%%%%%%%%%%%%%%%%%%%%%%%%%%%%%%%%%%%%%%%%%%%%%%%%%
\section{Review of experimental topologies and results} \label{searches}
The nature of the NLSP determines
the final state topologies to be studied in GMSB models. 
All the relevant searches according to the NLSP type and
lifetime are listed in Table~\ref{topo}.
\begin{table}[htb]
\begin{center}
\caption[]{\label{topo}{\small Final state topologies studied in the
different GMSB scenarios.}}
\vspace*{0.2cm}
\begin{tabular}{|l|l|l|l|l|}
\hline
 NLSP & Production & Decay mode & Decay length & Expected topology  \\
\hline
\hline
  &   &  & $\lambda \ll \ldet$ & Acoplanar photons  \\
 $\neu$ &  $\ee \to \neu \neu$ & $\neu \to \gamma \grav$  &  $\lambda \sim \ldet$ & Non-pointing photon  \\
 &  &  &  $\lambda \gg \ldet$ & [Indirect search] \\
\hline
\hline
 &  &  & $\lambda \ll \ldet$ & Acoplanar leptons  \\
 $\slep$ & $\ee \to \slep \slep$ & $\slep \to \ell \grav$   &  $\lambda \sim \ldet$ & Kinks and large impact parameters  \\
 &  &  &  $\lambda \gg \ldet$ & Heavy stable charged particles   \\
\hline
 &  & &  $\lambda \ll \ldet$ & Four leptons   \\
 $\slep$ &  $\ee \to \neu \neu$ & $\neu \to \ell \slep \to \ell \ell \grav$  &  $\lambda \sim \ldet$ & Four leptons with lifetime  \\
 &  &  &  $\lambda \gg \ldet$ & [Not covered here] \\
\hline
\end{tabular}
\end{center}
\end{table}

A detailed description of the selections optimised at 189\,GeV 
can be found in Ref.~\cite{gmsbpaper}. The same selections are applied
here with cut values suitably adjusted to take into account the beam
energy and luminosity increases. Only the four-lepton selections, detailed in
Section~\ref{cascade}, and the acoplanar lepton selection, recently
updated in Ref.~\cite{slepba}, have been modified.
%%%%%%%%%%%%%%%%%%%%%%%%%%%%%%%%%%%%%%%%%%%%%%%%%%%%%%%%%%%%%%%%%%%%%%
% Neutralino NLSP
%%%%%%%%%%%%%%%%%%%%%%%%%%%%%%%%%%%%%%%%%%%%%%%%%%%%%%%%%%%%%%%%%%%%%%
\subsection{Neutralino NLSP}\label{neu}
In the $\neu$ NLSP scenario, all supersymmetric decay chains will
terminate in $\neu \to \grav \gamma$. Searches for pair production
of neutralinos decaying promptly and neutralinos with intermediate
lifetime were described in Ref.~\cite{gmsbpaper} and updated in
Ref.~\cite{photons}. For short neutralino lifetimes,
the resulting experimental signature is a pair of energetic acoplanar 
photons. When the updated analysis is applied to the 189--209\,GeV data
sample, 
four candidate events are found with 4.9 events expected 
from background processes. 
For intermediate $\neu$ lifetimes, one neutralino may decay before reaching
the electromagnetic calorimeter, while the other decays outside
the detector. This scenario results  in a topology where the only visible
energy originates from a single photon which does not 
point to the interaction vertex. Two non-pointing photon events are
found in the data sample and 1.0 are expected. 
Systematic errors have been evaluated and included in the results as in 
Ref.~\cite{gmsbpaper}.

For long $\neu$ NLSP decay lengths  
($\lambda \gg \ldet$) the neutralino becomes 
invisible and only indirect exclusions are possible. 
The relationship between the $\neu$ mass and the 
slepton and chargino masses can be exploited to put indirect limits on
the $\neu$ mass using the ALEPH results on 
slepton~\cite{slepba} and chargino~\cite{charg01} searches performed within
the SUGRA framework.

\begin{figure}[htb]
\begin{center}
\epsfig{file=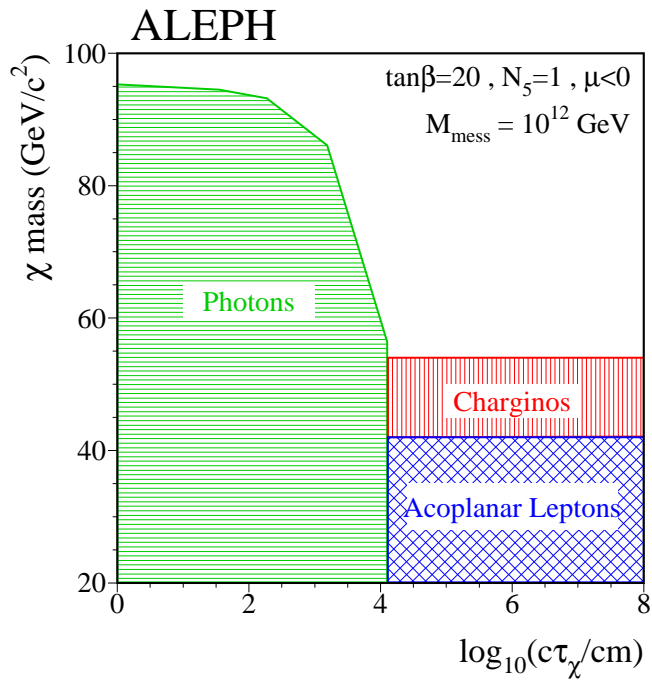,width=10cm}
\caption[]{\label{chilife}{\small Excluded neutralino
mass at 95$\%$ confidence level as a function of its lifetime in the  
neutralino NLSP scenario.
The excluded areas are obtained by the scan described in Section~\ref{scan} 
for negative $\mu$, $N_5 = 1$, $M_{\mathrm{mess}} = 10^{12}\gevcc$, $\tanb
= 20$ and any $\Lambda$. 
The short and medium lifetime cases, when at least one $\neu$ decays 
inside the detector, are covered by the
acoplanar photons and non-pointing photon searches. For long-lived
neutralinos the gravity mediated searches for charginos and sleptons are used.}}
\end{center}
\end{figure}

The combination of all these analyses allows the exclusion of
neutralino masses as a function of the neutralino decay length.
An example for a particular region in the GMSB parameter space is shown
in Fig.~\ref{chilife}.

%%%%%%%%%%%%%%%%%%%%%%%%%%%%%%%%%%%%%%%%%%%%%%%%%%%%%%%%%%%%%%%%%%%%%%
% Slepton NLSP
%%%%%%%%%%%%%%%%%%%%%%%%%%%%%%%%%%%%%%%%%%%%%%%%%%%%%%%%%%%%%%%%%%%%%%
\subsection{Slepton NLSP direct decay}\label{slepdirect}
In the case of a slepton NLSP, the pair-production process $\ee \to
\slep \slep$ is expected to be the main experimental 
signature. The signal final state topology depends strongly
on the slepton lifetime. Four different analyses are performed, each 
corresponding to a specific range of mean decay length. These
searches were described in Ref.~\cite{oldcasc} and updated 
in Ref.~\cite{gmsbpaper}.

If the $\slep$ has a short
decay length, of the order of a few mm or less, the
final state topology will be a pair of acoplanar leptons and missing
energy, carried away by the two gravitinos. This final state is studied in
gravity mediated models~\cite{slepba}, 
where a neutralino of almost zero mass plays
the role of the gravitino in GMSB. Therefore, the results obtained for
a massless neutralino are used. 

Sleptons with intermediate lifetime, which decay inside the
detector, may show two possible signatures: large impact parameter and
kinked tracks. If the slepton decays before the TPC,
between $\sim$1\,cm and 40\,cm, the slepton track will not be reconstructed and
the final lepton track will have a large impact parameter.
If the slepton decays within the TPC
volume, the signature is then characterised by a kinked track. 
Two different selections are therefore applied to the
intermediate slepton lifetime case.  One event
is found with 1.1 events expected from SM processes.

Finally, long-lived sleptons can be detected from their
anomalous specific ionisation in the TPC. The search for heavy stable
charged particles selected one candidate event, 
while 2.3 are expected from background processes. 

\begin{figure}[!ht]
\begin{center}
\epsfig{file=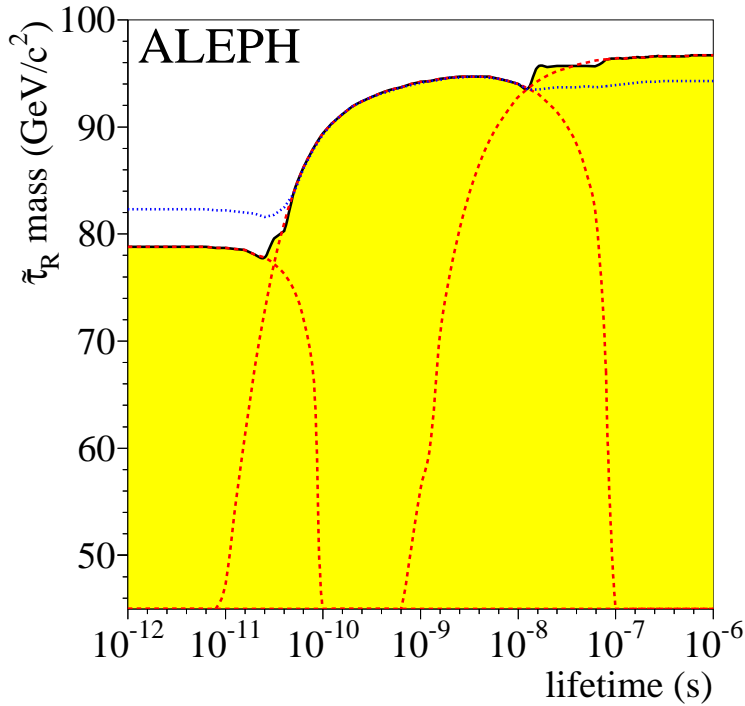,width=10cm}
\caption[]{\label{slepcomb}{\small Excluded $\stau_{\mathrm{R}}$ mass at 95$\%$
C.L. as a function of its lifetime (shaded area) from direct searches. 
Dashed curves give the limits from
the different topologies. The search for acoplanar leptons covers the 
case of small lifetimes, 
searches for tracks with large impact parameter and for kinks are
used in the intermediate range, whereas for very large lifetimes a 
search for heavy stable charged particles is performed. 
The dotted curve gives the expected limit.}}
\end{center}
\end{figure}

The effect of systematic uncertainties on kinematic cuts has been studied
as in Ref.~\cite{gmsbpaper} and limits are derived reducing the efficiency
by a total systematic error of 5$\%$.
When all four independent selections are combined, 
the 95$\%$ confidence level lower limits on the
right-slepton masses, independent of lifetime, are set at 
83, 88 and 77$\gevcc$ for selectron, smuon and stau, respectively.
The selectron mass limit is obtained neglecting the
$t$-channel exchange contribution to the production cross section.
The stau mass limit as a function of lifetime is plotted in
Fig.~\ref{slepcomb}.  

%%%%%%%%%%%%%%%%%%%%%%%%%%%%%%%%%%%%%%%%%%%%%%%%%%%%%%%%%%%%%%%%%%%%%%
% Slepton cascade
%%%%%%%%%%%%%%%%%%%%%%%%%%%%%%%%%%%%%%%%%%%%%%%%%%%%%%%%%%%%%%%%%%%%%%%%
\subsection{Slepton NLSP cascade decays} \label{cascade}
In the $\slep$ NLSP scenario, if the neutralino pair production 
is kinematically accessible, the process
$\ee \to \neu \neu \to \ell \slep \: \ell \slep \to \ell \ell 
\grav \: \ell \ell \grav$ may provide a very distinctive discovery signal.
This cascade decay can increase the sensitivity 
to GMSB signatures in some region of the parameter space.
This process benefits from quite a large cross section 
(the $\neu$ is expected to be mainly bino and the $\sel_{\mathrm{R}}$ is 
expected to be light) and from a clear experimental signature. Four 
leptons are produced in the final state (two could be soft if 
the $\neu$-$\slep$ mass difference is small) and in half of the cases 
the two most energetic leptons have the same charge ($\neu$ are Majorana
particles).

Depending on the flavour of the slepton in $\chi \rightarrow
\ell \tilde{\ell}$ decays, there are six
possible topologies, labelled $\sel\sel$, $\smu\smu$,
$\stau\stau$, $\sel\smu$, $\sel\stau$ and
$\smu\stau$ in the case of slepton co-NLSP. In the case of
stau NLSP only the $\stau\stau$ topology is relevant.

%%%%%%%%%%%%%%%%%%%%%%%%%%%%%%%%%%%%%%%%%%%%%%%%%%%%%%%%%%%%%%%%%%%%%%
% Eva's Search
\subsubsection{Prompt decays} 
Searches for the cascade topologies with negligible lifetime 
have already been performed by ALEPH at a
centre-of-mass energy of 189\,GeV~\cite{gmsbpaper}. 
The selection cuts are revisited in this paper to
improve the signal efficiencies and account for the increase 
in centre-of-mass energy and luminosity. 
The revised cuts described in Appendix~A are also applied to the data 
collected at 189\,GeV and improve the signal efficiency up to 10$\%$.

The main systematic uncertainties on the background and signal expectations
come from the number of generated events in the simulated samples (up to
$4\%$). A total systematic uncertainty of $2\%$ is evaluated for the
variables involved in the selection. To derive the results, 
the selection efficiency has been reduced by 4$\%$. 

The efficiency of the cuts on the signal samples was found to be in
the range 65--85$\%$ for $\neu$-$\slep$ mass differences greater
than 3$\gevcc$ in the case of final states not involving tau leptons
and in the range 40--65$\%$ for $\neu$-$\slep$ mass differences
greater than 5$\gevcc$ in the case of final states with tau leptons. 
The selection efficiencies for the six topologies are shown in Fig.~\ref{cascade_eff}a.

The numbers of background events expected from Standard Model processes and
of events observed in the data
are given in Table~\ref{obscascade}. The largest background contributions are
from WW and $\tau\tau(\gamma)$ events. 

\begin{table}[tb]
\center
\caption{\label{obscascade} \small Expected Standard Model background and selected candidates
for the various cascade topologies in the case of negligible slepton
lifetime.}
\vspace*{0.2cm}
\begin{tabular}{|c|c|c|c|c|c|c|c|c|c|c|c|c|} \hline
Energy  & \multicolumn{2}{c|}{$\sel\sel$} & \multicolumn{2}{c|}{$\smu\smu$} & \multicolumn{2}{c|}{$\sel\smu$} & \multicolumn{2}{c|}{$\stau\stau$} & \multicolumn{2}{c|}{$\sel\stau$} & \multicolumn{2}{c|}{$\smu\stau$} \\
%\cline{1-2}
\cline{2-13}
(GeV)   & exp & obs & exp & obs & exp & obs & exp & obs & exp & obs & exp & obs \\
\hline
$188.6$ & 1.33 & 2 & 0.12 & 1 & 0.98 & 1 & 5.23 & 4 & 1.34 & 2 & 1.77 & 0 \\
$191.6$ & 0.31 & 1 & 0.02 & 0 & 0.29 & 0 & 1.17 & 1 & 0.33 & 1 & 0.29 & 1 \\
$195.5$ & 1.05 & 2 & 0.07 & 0 & 0.41 & 0 & 1.84 & 7 & 0.72 & 0 & 0.61 & 0 \\
$199.5$ & 0.88 & 1 & 0.04 & 0 & 0.69 & 0 & 2.19 & 2 & 0.85 & 2 & 0.86 & 1 \\
$201.6$ & 0.27 & 0 & 0.03 & 0 & 0.17 & 1 & 0.97 & 3 & 0.30 & 0 & 0.33 & 0 \\
$205.0$ & 0.51 & 0 & 0.07 & 0 & 0.41 & 0 & 1.96 & 1 & 0.72 & 0 & 0.62 & 1 \\ 
$206.7$ & 0.80 & 0 & 0.09 & 0 & 0.70 & 0 & 3.13 & 4 & 1.26 & 3 & 1.24 & 2 \\ 
\hline
Total &  5.15 & 6 & 0.44 & 1 & 3.65 & 2 & 16.49 & 22 & 5.52 & 8 & 5.72 & 5 \\
\hline
\end{tabular}
\end{table}
\normalsize

%%%%%%%%%%%%%%%%%%%%%%%%%%%%%%%%%%%%%%%%%%%%%%%%%%%%%%%%%%%%%%%%%%%%%%
% Luke's Search
\subsubsection{Short and long decays}  \label{luke}
A new analysis has been developed for the case of observable slepton 
decay length, based on the experimental signatures of leptons
with large impact parameter and track kinks. 
Searches were developed for each of the six channels, consisting of loose 
cuts on global event variables, which cause very
little signal rejection, and of more stringent cuts on individual track
properties. The latter are intended to select the tracks that come from the
decay of the long-lived sleptons. For each topology, two independent
selections were designed to ensure good sensitivity to the signal both for
short ($\sim$1\,cm) and long ($\sim$1\,m) decay lengths. 
Further details are given in Appendix B. 

Because the searches in all topologies focus primarily on
selecting tracks with large $d_0$, there is a large overlap in their
acceptances, and so candidate events tend to be selected  more than once.
 The ``efficiency'' (the probability for a signal event to be selected
at least once) reaches a maximum of $\sim$80$\%$ for
the $\stau\stau$ channel and $>$90$\%$ for other channels at a slepton
decay length of around 10\,cm. An efficiency $>$10$\%$ is maintained
for slepton decay lengths from $\sim$1\,mm to $\sim$10\,m for all
channels.

\begin{figure}[htb]
\begin{center}
\epsfig{file=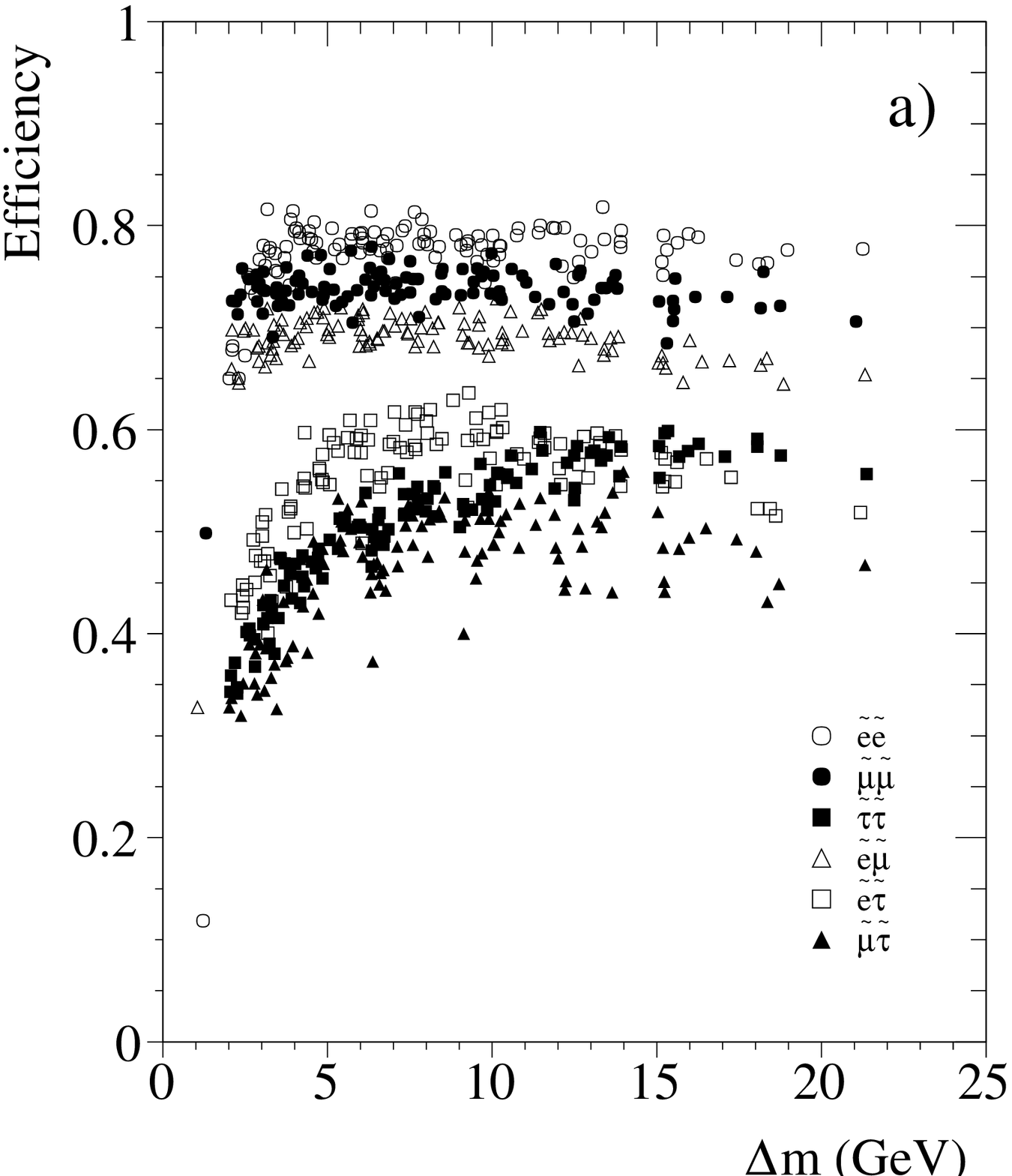,height=8.5cm}
\epsfig{file=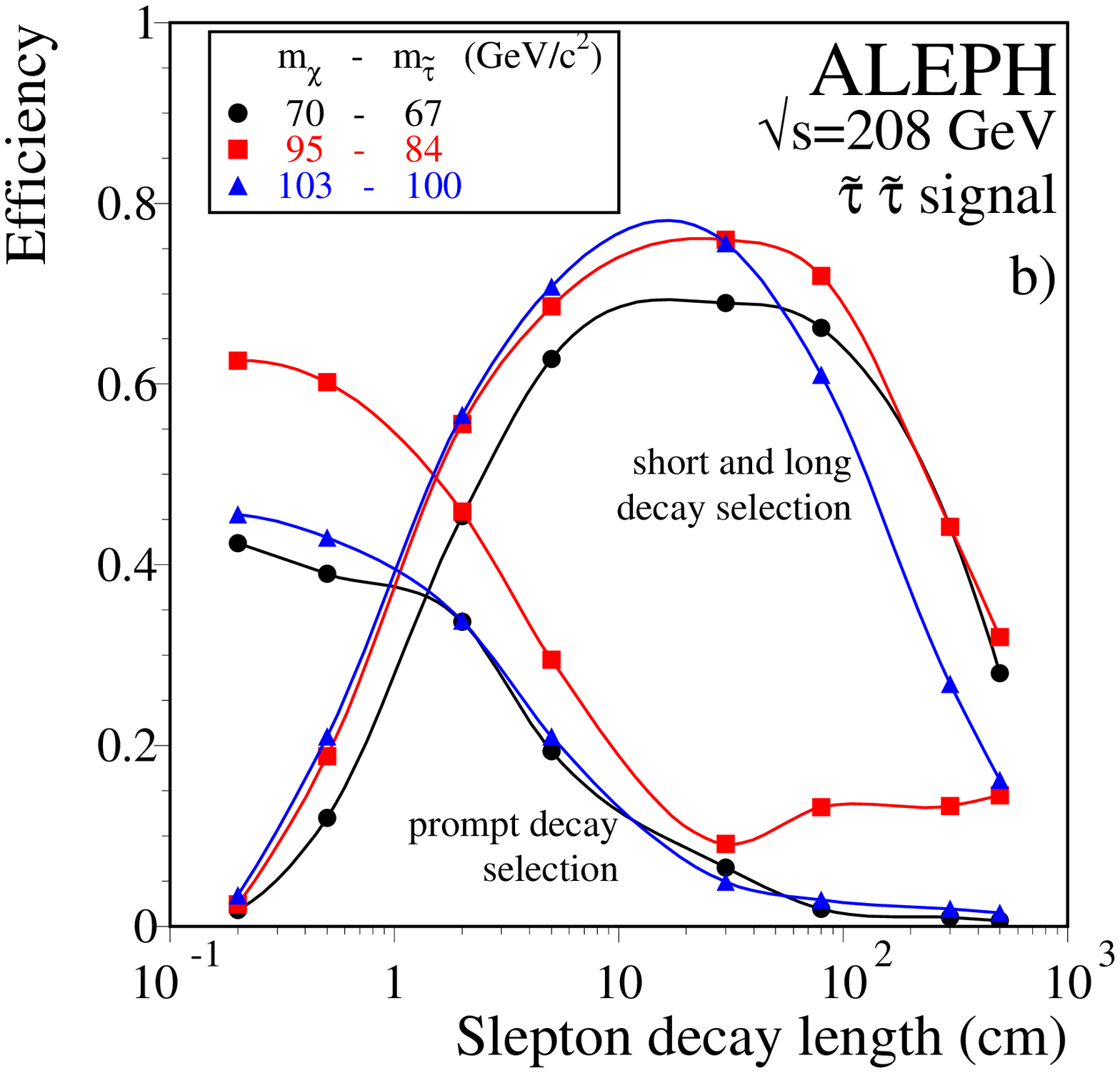,height=9.2cm}
\caption{\label{cascade_eff}{\small a) Selection efficiencies for the
six different event topologies versus $\Delta m$ =
$m_{\neu}$-$m_{\slep}$ in the zero slepton lifetime case. The spread
of points observed for a given topology is due to different 
values of neutralino and slepton masses.
b) Probability for a signal stau-pair event
to be selected by at least one of the six topological searches versus
slepton decay length. 
The set of curves with higher efficiency in the 0.1\,cm area
corresponds to the prompt decay selection. Those peaking at $\sim$10\,cm correspond
to the short and long decay length selection. Different lines correspond to
different points in the ($m_{\neu}$,$m_{\stau}$) space.}}
\end{center}
\end{figure}

The numbers of background events expected to
be selected in at least one topology and the corresponding numbers of
observed events for each LEP energy and each selection are given in
Table~\ref{count}.
The efficiency dependence on the slepton decay length is shown in
Fig.~\ref{cascade_eff}b, for the prompt decay and the short and long decay
length selections.

\begin{table}[tb]
\center
\caption{\label{count} \small Summary of results for the case of observable slepton
decay length in cascade decays: the numbers of observed candidate
events passing at least one topology and of expected
background events for both decay length selections.}
\vspace*{0.2cm}
\begin{tabular}{|c|c|c|c|c|} \hline
Energy & \multicolumn{2}{c|}{Short decay length} & 
\multicolumn{2}{c|}{Long decay length} \\ \cline{2-5}
(GeV)  & expected & observed & expected & observed  \\ 
\hline
188.6  & 1.39 & 0 & 0.51 & 1 \\
191.6  & 0.24 & 2 & 0.08 & 1 \\
195.5  & 0.66 & 1 & 0.20 & 1 \\
199.5  & 0.73 & 0 & 0.20 & 0 \\
201.6  & 0.38 & 0 & 0.10 & 0 \\
205.0  & 0.69 & 0 & 0.16 & 0 \\
206.7  & 1.16 & 2 & 0.26 & 1 \\ \hline
Total  & 5.25 & 5 & 1.51 & 4 \\ \hline
\end{tabular}
\end{table}

%%%%%%%%%%%%%%%%%%%%%%%%%%%%%%%%%%%%%%%%%%%%%%%%%%%%%%%%%%%%%%%%%%%%%%
% Scan 
%%%%%%%%%%%%%%%%%%%%%%%%%%%%%%%%%%%%%%%%%%%%%%%%%%%%%%%%%%%%%%%%%%%%%%
\section{Interpretation of the results in the minimal GMSB model} \label{scan}
A scan over the parameters of a minimal GMSB model has been performed
to study the impact of the different searches. 
The aim of this scan is to understand which topologies contribute to
exclude regions in the parameter space and to be able to set a lower
limit on the mass of the NLSP and on the universal mass scale
$\Lambda$, independently of the NLSP lifetime 
(i.e., for all gravitino masses).

The parameters needed to specify a minimal GMSB model~\cite{theory} are
\begin{itemize}
\item $\Lambda$, the universal mass scale of SUSY particles; 
\item $N_5$, the number of messenger pairs; 
\item $M_{\mathrm{mess}}$, the common messenger mass scale;
\item $\tan\beta$, the ratio between the vacuum expectation values of the
two Higgs doublets; 
\item sign($\mu$), where $\mu$ is the higgsino  mass
parameter; and 
\item $\sqrt{F}$, the SUSY breaking scale in the messenger sector, related to the
gravitino  mass by
\end{itemize}
\begin{equation} \label{mg-rootf}
 m_{\grav} = \frac{F}{\sqrt{3}~M_{\mathrm{Planck}}} = 2.4 \left (
\frac{\sqrt{F}}{100\,\mathrm{TeV}} \right )^2\,\mathrm{eV},
\end{equation}
where $M_{\mathrm{Planck}}=2.4 \times 10^{18}\gevcc$ is the reduced
Planck mass.

The ranges of the scan are listed in Table~\ref{range}. 
The six parameters listed determine the properties of supersymmetric
particles characteristic of GMSB models. At each point in the scan,
the {\tt ISAJET 7.51} program~\cite{isajet} was used to calculate SUSY
masses, cross sections, branching ratios and lifetimes, then taken into
account to evaluate whether a point is excluded by any of the searches. 
In total, over 2.3 million points in the minimal GMSB parameter space have
been tested.   

\begin{table}[tb]
\begin{center}
\caption[]{\label{range}{\small Minimal set of parameters and their ranges of
variation in the scan.}}
\vspace*{0.2cm}
\begin{tabular}{|c|c|c|}
\hline
 Parameter  & Lower limit  & Upper limit  \\
\hline
\hline
$M_{\mathrm{mess}}$   & 10$^4$ GeV    &  10$^{12}$ GeV              \\
$m_{\grav}$  	      & 10$^{-1}$ eV  &  10$^{5}$ eV                \\
$\Lambda$             & 10$^3$ GeV    &  min($\sqrt{F}$, $M_{\mathrm{mess}}$) \\
$\tanb$               & 1.5           &  40                         \\
$N_5$                 & 1             &  5                          \\
sign($\mu$)           & $-$           &  +                          \\
\hline
\end{tabular}
\end{center}
\end{table}

In addition to the analyses described in this paper, other searches
were used to set exclusion areas in the scan:
the SUGRA chargino~\cite{charg01} and 
slepton~\cite{slepba} searches to cover the case of $\neu$ NLSP
with a long-lived neutralino and LEP1 results~\cite{stauLEP1,EW},
used here to exclude very low NLSP masses.
In addition, for each set of GMSB parameters the Higgs boson masses and
couplings were computed. The results from Ref.~\cite{aleph_higgs} were used
to extend the GMSB exclusion domain.
%%%%%%%%%%%%%%%%%%%%%%%%%%%%%%%%%%%%%%%%%%%%%%%%%%%%%%%%%%%%%%%%%%%%%%
% Limit on m_NLSP
\subsection{Lower limit on the NLSP mass}
As reported in Section~\ref{searches}, no significant deviation from the SM
expectation was observed. A lower limit on the $\stau_{\mathrm{1}}$ mass of 77$\gevcc$ is set,
independently of its lifetime. This limit is valid in the stau NLSP case
over the full scan range. It is reduced to 72$\gevcc$ in the $\neu$ NLSP
scenario. 

\begin{figure}[!ht]
\begin{center}
\epsfig{file=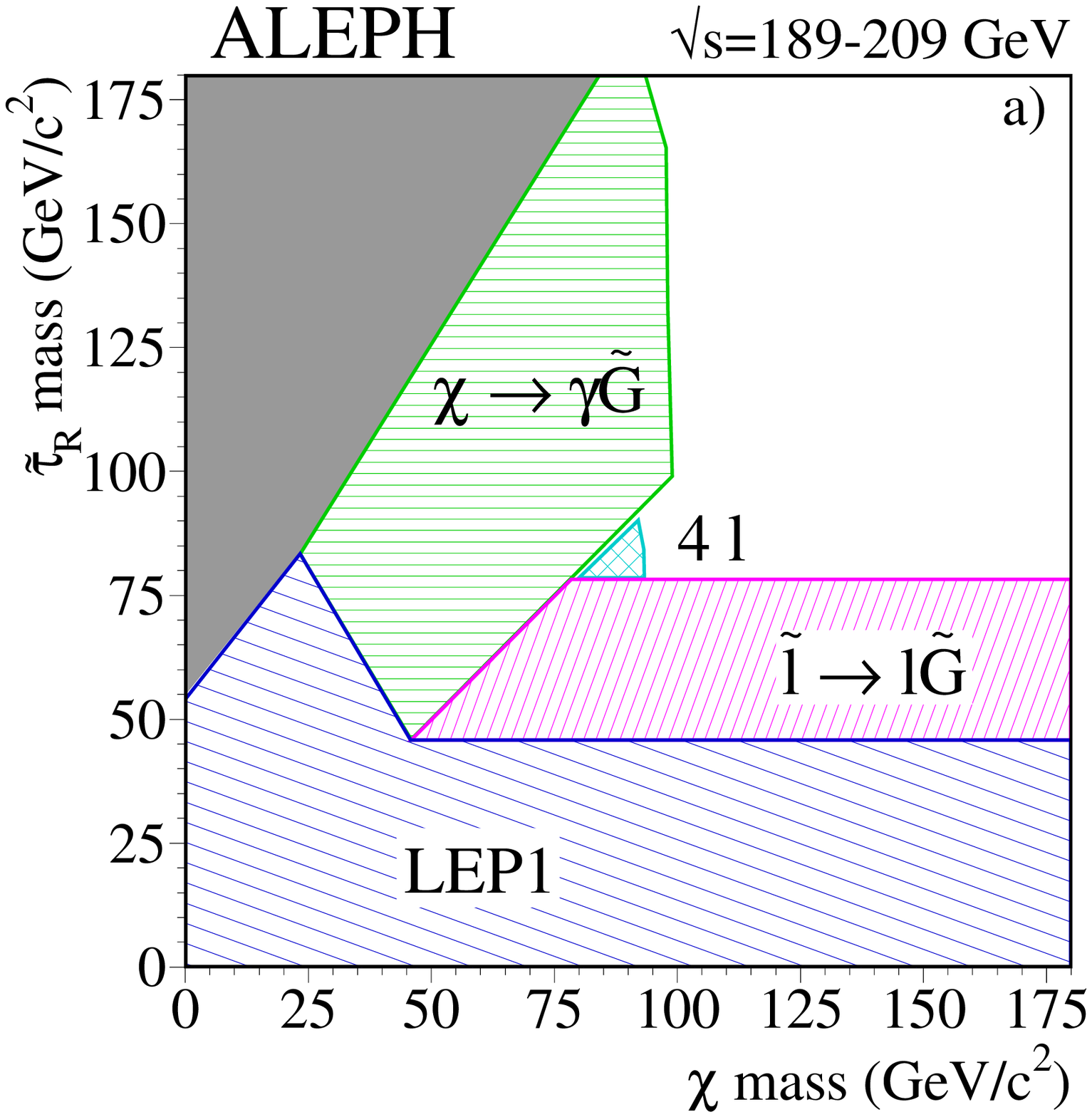,width=8.4cm}
\epsfig{file=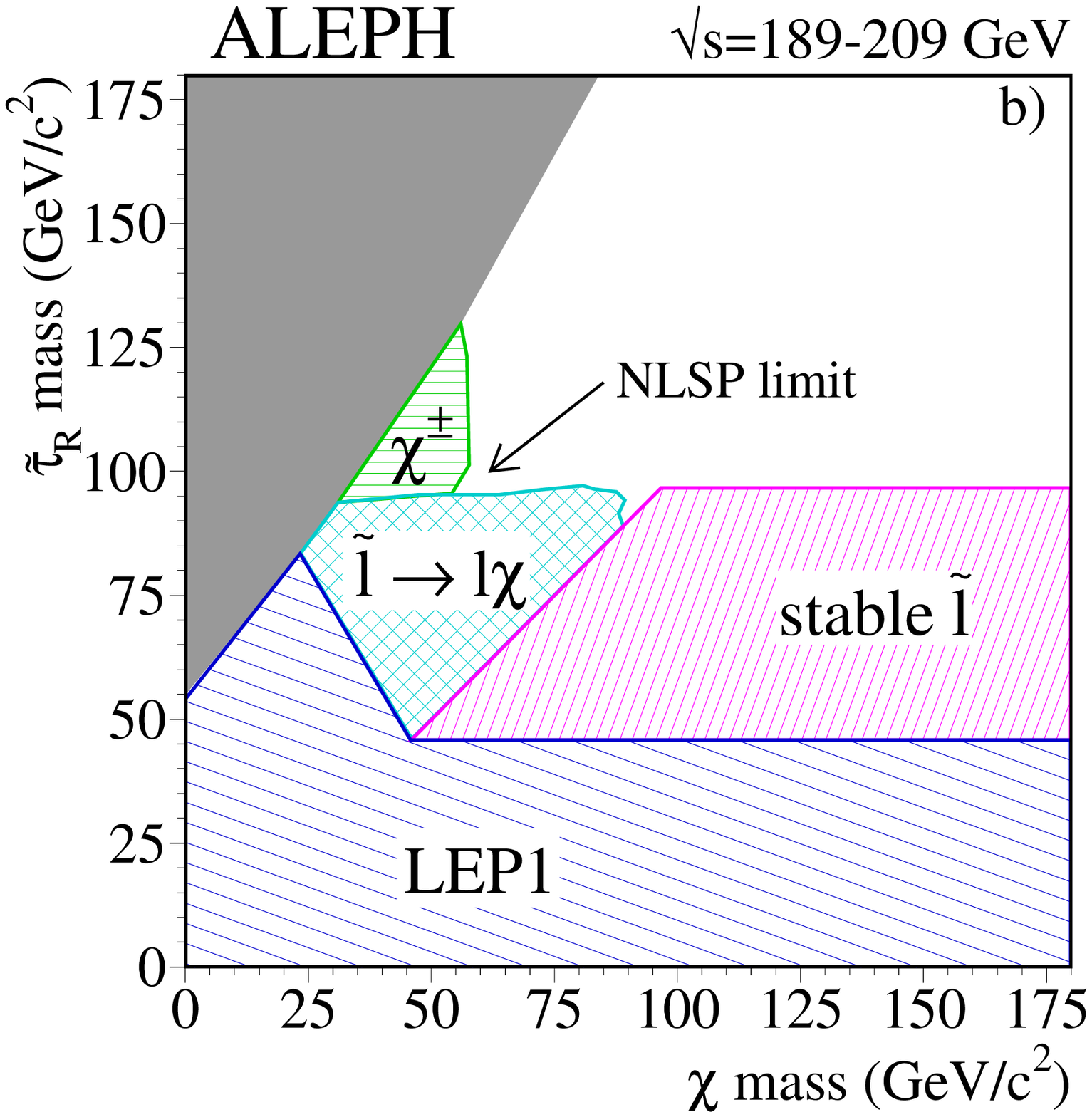,width=8.4cm}
\caption[]{\label{neustau}{\small Regions excluded by the different analyses
described in the text at 95$\%$ confidence level in the 
($m_{\neu},m_{\stau}$) plane, for a) short NLSP lifetimes ($m_{\grav} \leq
10\evcc$) and b) long NLSP lifetimes ($m_{\grav} \geq 1\kevcc$). 
Points in the dark region are not accessible to the scan. 
The absolute NLSP mass limit is set at 54$\gevcc$ in b) by
the intersection of chargino and slepton searches.}}
\end{center}
\end{figure}

The interplay of the different searches in the
($m_{\neu},m_{\stau}$) plane is shown in Fig.~\ref{neustau}. For short
NLSP lifetimes (Fig.~\ref{neustau}a) the multi-lepton search is
able to exclude $m_{\neu}$ up to 92$\gevcc$ in the slepton NLSP case,
extending the acoplanar lepton search. 
In Fig.~\ref{neustau}b the case of long NLSP
lifetimes is presented. Because they rely on results from indirect
constraints, limits in the long neutralino lifetime case
are less constraining than those obtained with short neutralino
lifetime searches. The absolute lower limit on the NLSP
mass of 54$\gevcc$, determined by the chargino and
slepton searches, is visible in Fig.~\ref{neustau}b. 
This point is found at $N_5 = 1$, $\tanb = 3$, $\Lambda = 39\tevcc$,
$M_{\mathrm{mess}} = 10^{10}\gevcc$ and $m_{\grav} = 10^{5}\evcc$, where
the neutralino is the NLSP with the $\slep$ masses around 96$\gevcc$ and
all other supersymmetric particles above threshold.

\begin{figure}[htb]
\begin{center}
\epsfig{file=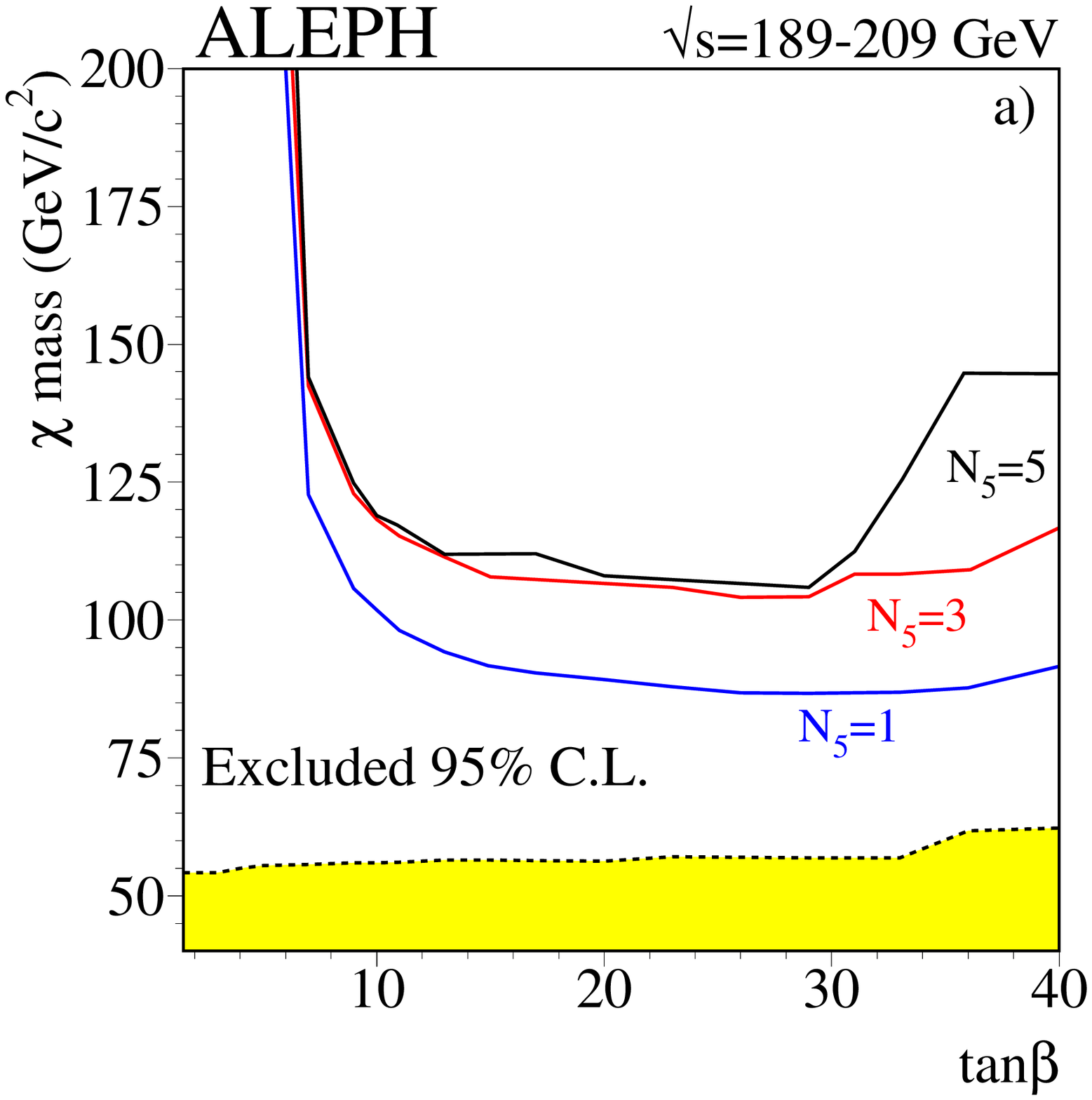,width=8.4cm}
\epsfig{file=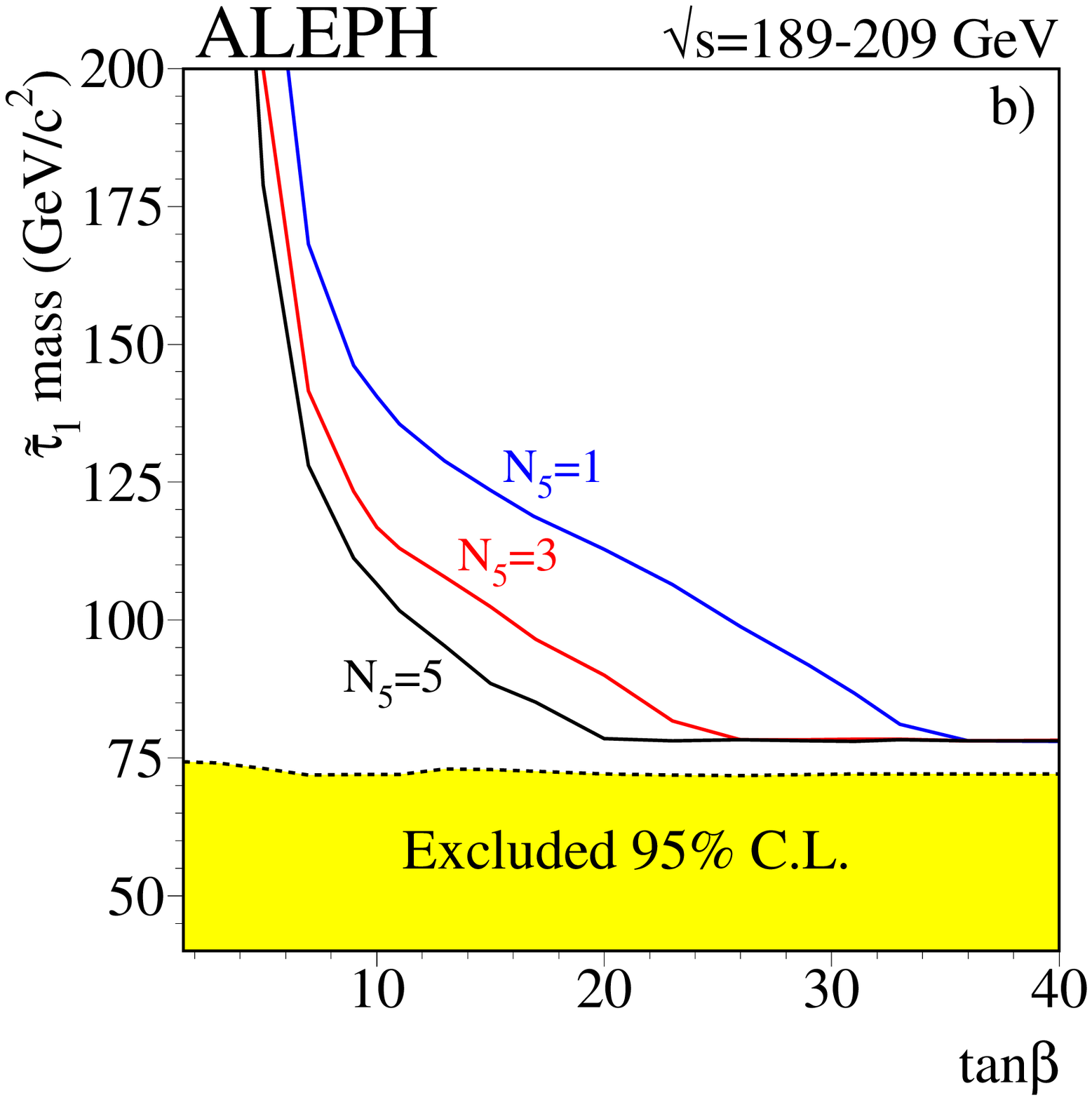,width=8.4cm}
\caption[]{\label{mnslphiggs}{\small 
Lower limits on the masses of a) $\neu$ and b) $\stau_1$ as a function
of $\tanb$, for different values of $N_5$, as set by the Higgs boson
searches. The shaded area represents the minimum excluded area, for any $N_5$, 
as derived from GMSB searches alone.}}
\end{center}
\end{figure}

The impact of the neutral Higgs boson searches on the neutralino and stau 
mass limits is shown in Fig.~\ref{mnslphiggs} as a function of $\tanb$. 
The NLSP absolute mass limit is 77$\gevcc$ obtained for large $\tanb$ and
in the stau NLSP scenario.  

%%%%%%%%%%%%%%%%%%%%%%%%%%%%%%%%%%%%%%%%%%%%%%%%%%%%%%%%%%%%%%%%%%%%%%
% Limit on Lambda
\subsection{Lower limit on $\Lambda$}
The parameter $\Lambda$ represents the energy scale at which the
messenger particles couple to the visible sector and hence fixes
the universal mass scale of SUSY particles. At the
$M_{\mathrm{mess}}$ energy, gaugino masses scale like $N_5\Lambda$, 
while scalar masses squared scale like $N_5\Lambda^2$. The masses at the
electroweak scale are calculated by means of the renormalization group
equations. 
Once the limit for the NLSP mass has been found,
the limit on $\Lambda$ as a function of $N_5$ can thus be derived.

\begin{figure}[tb]
\begin{center}
\epsfig{file=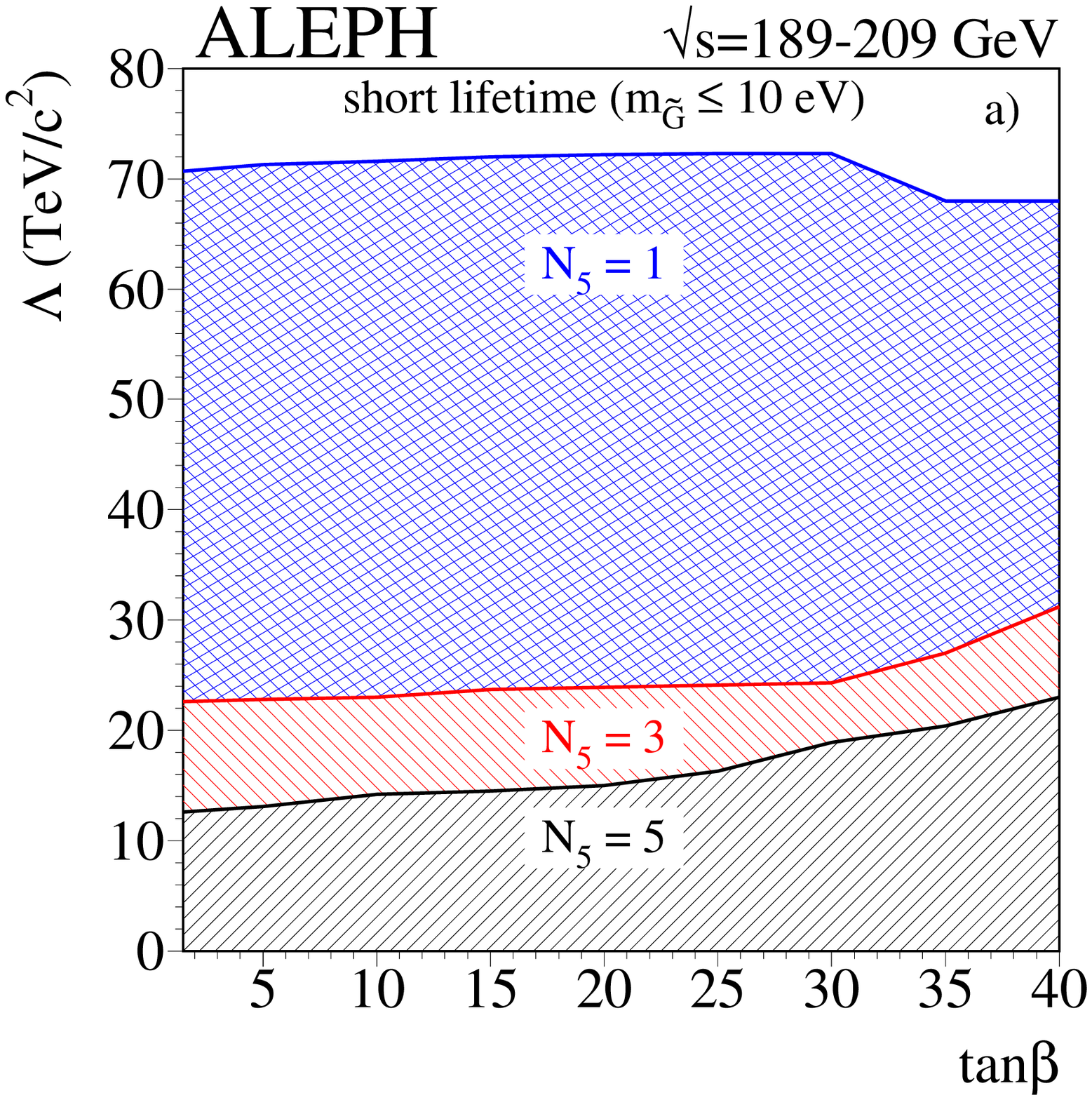,width=7.4cm}
\epsfig{file=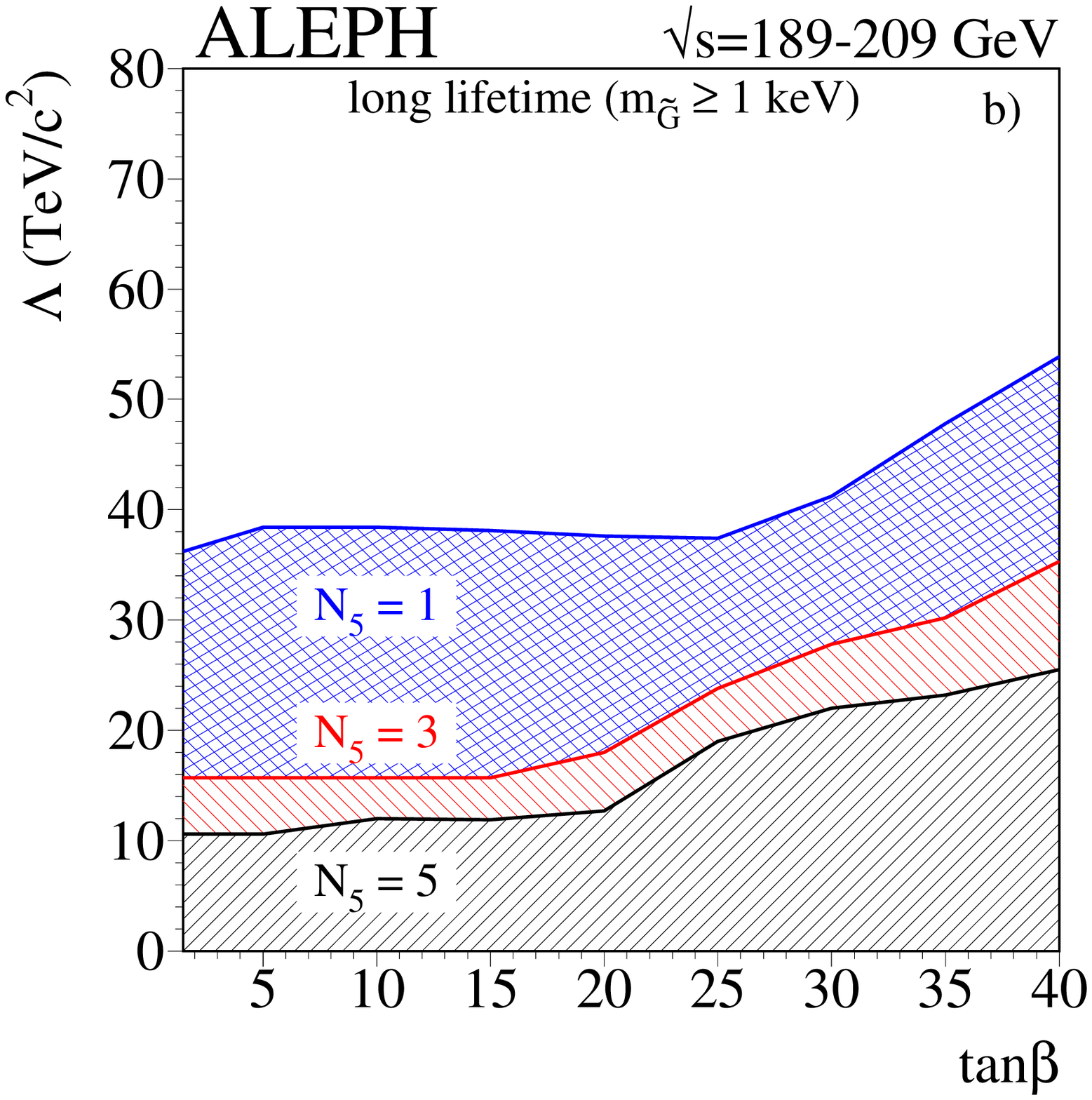,width=7.4cm}
\epsfig{file=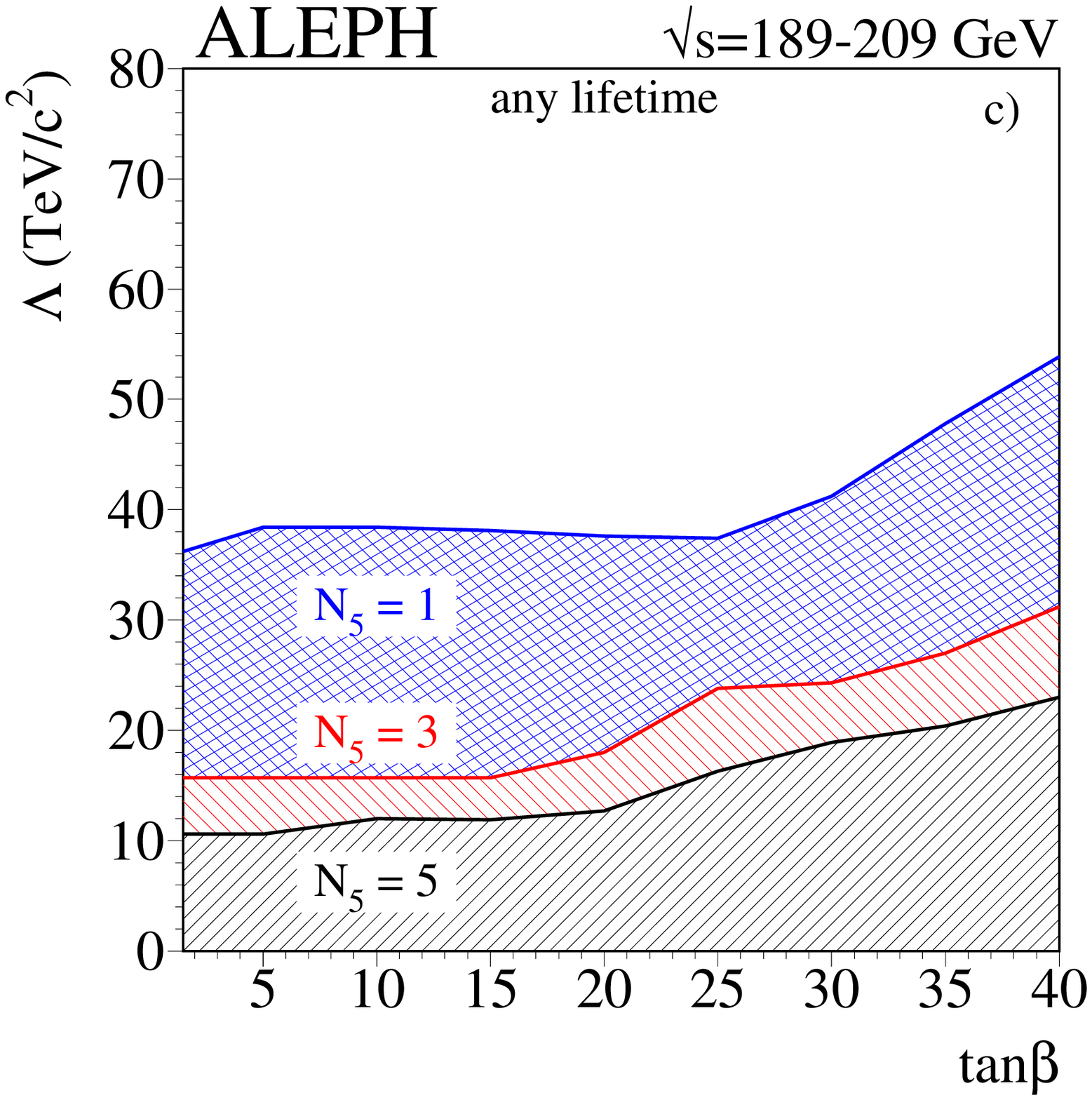,width=7.4cm}
\epsfig{file=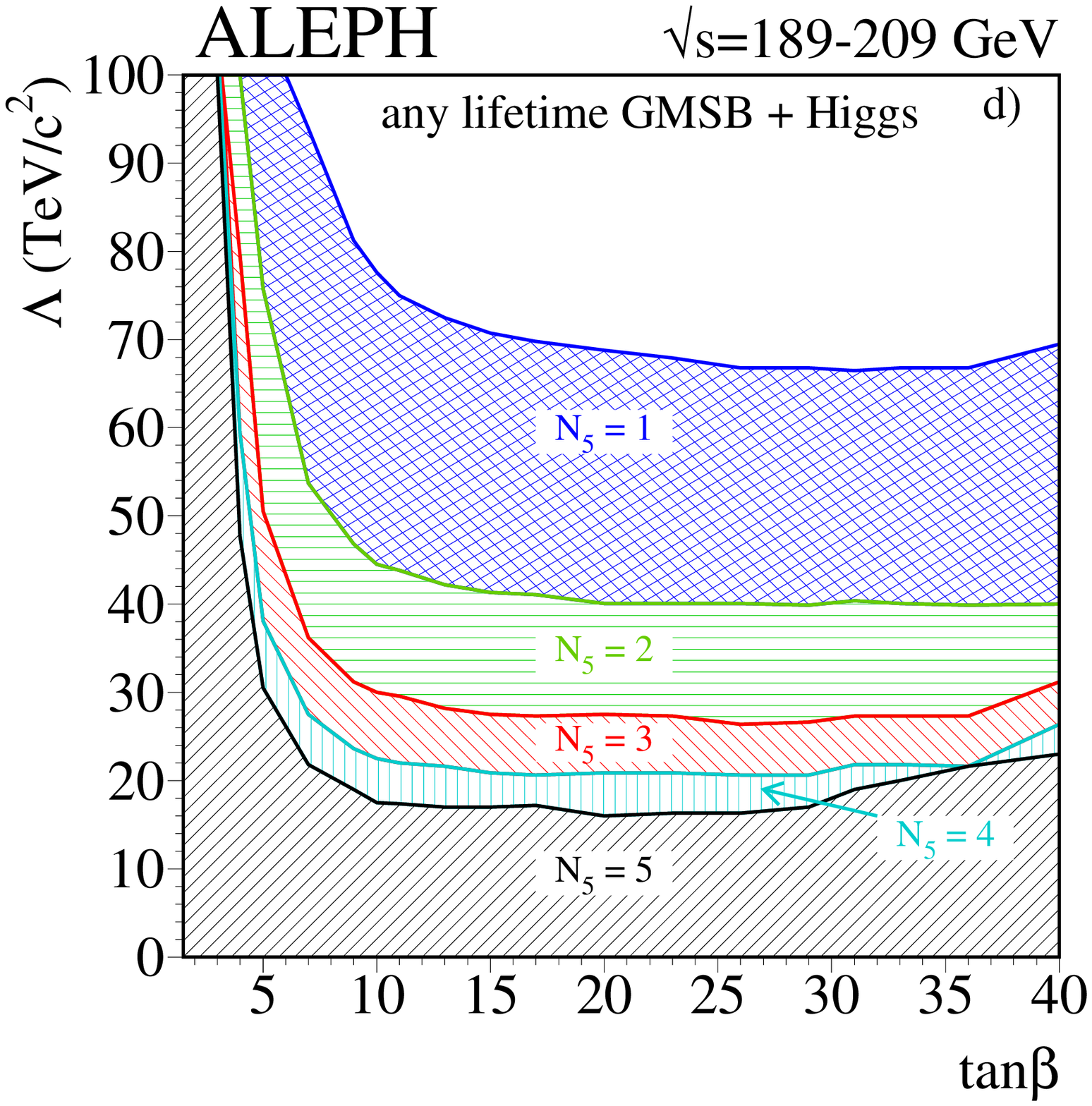,width=7.4cm}
\caption[]{\label{latan}{\small Region excluded at 95$\%$ C.L. in the 
($\Lambda,\tanb$) plane for a) short, b) long and c) any NLSP lifetime. 
The impact of the Higgs search is included in d). 
Values of $\tanb$ less than 3 are excluded for large $N_5$ at
any mass parameter $\Lambda$, while $\tanb$ up to 6 can be excluded for
$N_5 = 1$.}}
\end{center}
\end{figure}

The excluded values for the parameter $\Lambda$ as a function of $\tanb$
are shown in Fig.~\ref{latan} for different values of $N_5$. 
The absolute limit for $\Lambda$ is set at around
10$\tevcc$. This limit is set at $N_5 = 5$, $\tan\beta = 1.5$, 
$M_{\mathrm{mess}} = 10^{12}\gevcc$ and a large gravitino mass (stable NLSP). 
The neutralino is the NLSP with a mass of 73$\gevcc$, slepton masses are
around 76$\gevcc$ and all other sparticles are above threshold. 

The impact of the neutral Higgs boson searches~\cite{aleph_higgs} is shown
in the ($\Lambda,\tanb$) plane in Fig.~\ref{latan}d. These results strongly
constrain the allowed values of $\Lambda$ for small $\tanb$. 
For example, for $N_5 = 1$, $\Lambda$ up to 67$\tevcc$
and $\tanb$ up to 6 are excluded. 

\begin{figure}[!htb]
\begin{center}
\label{Lambda}
\epsfig{file=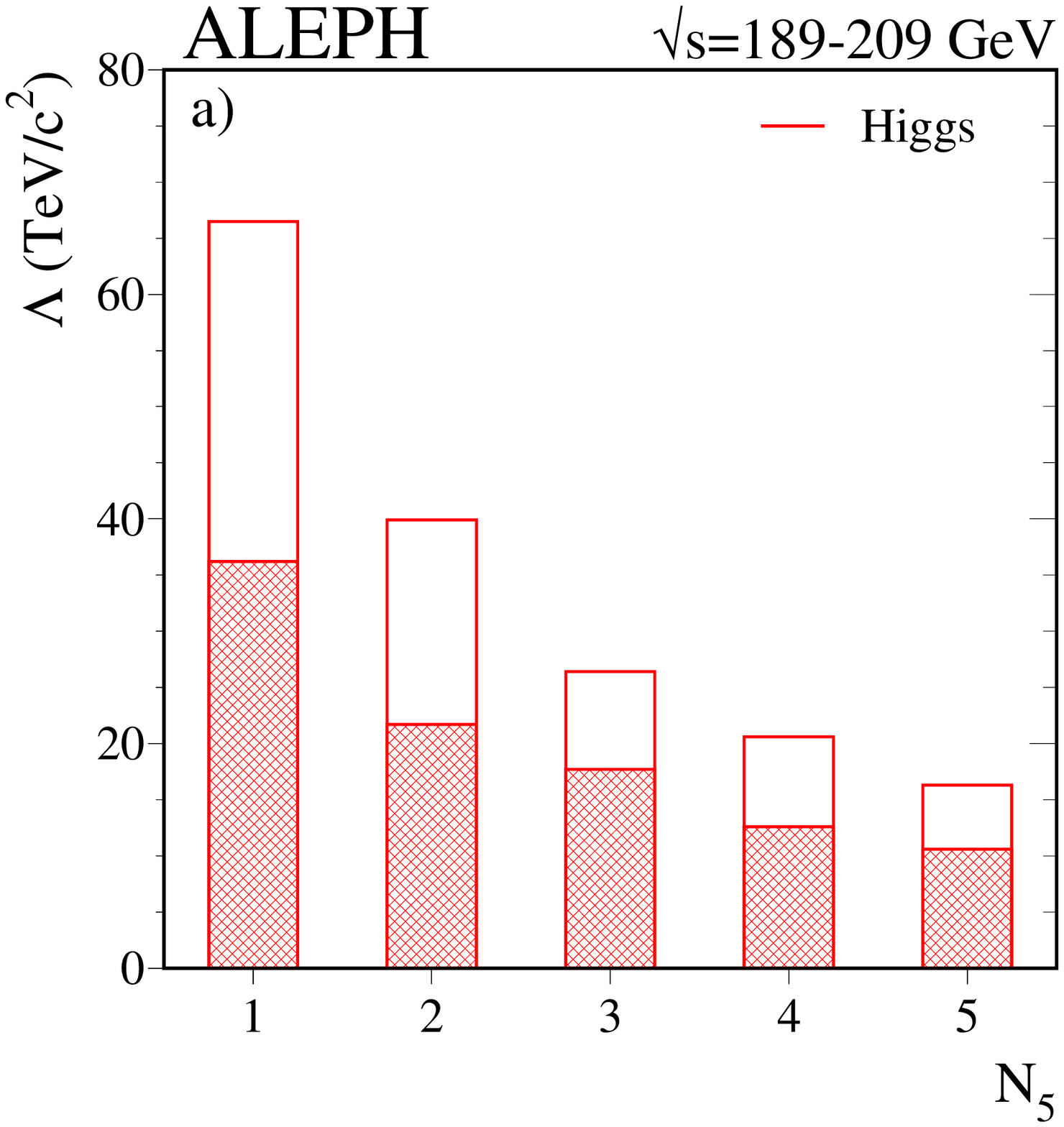,width=8cm}
\label{Mgrav}
\epsfig{file=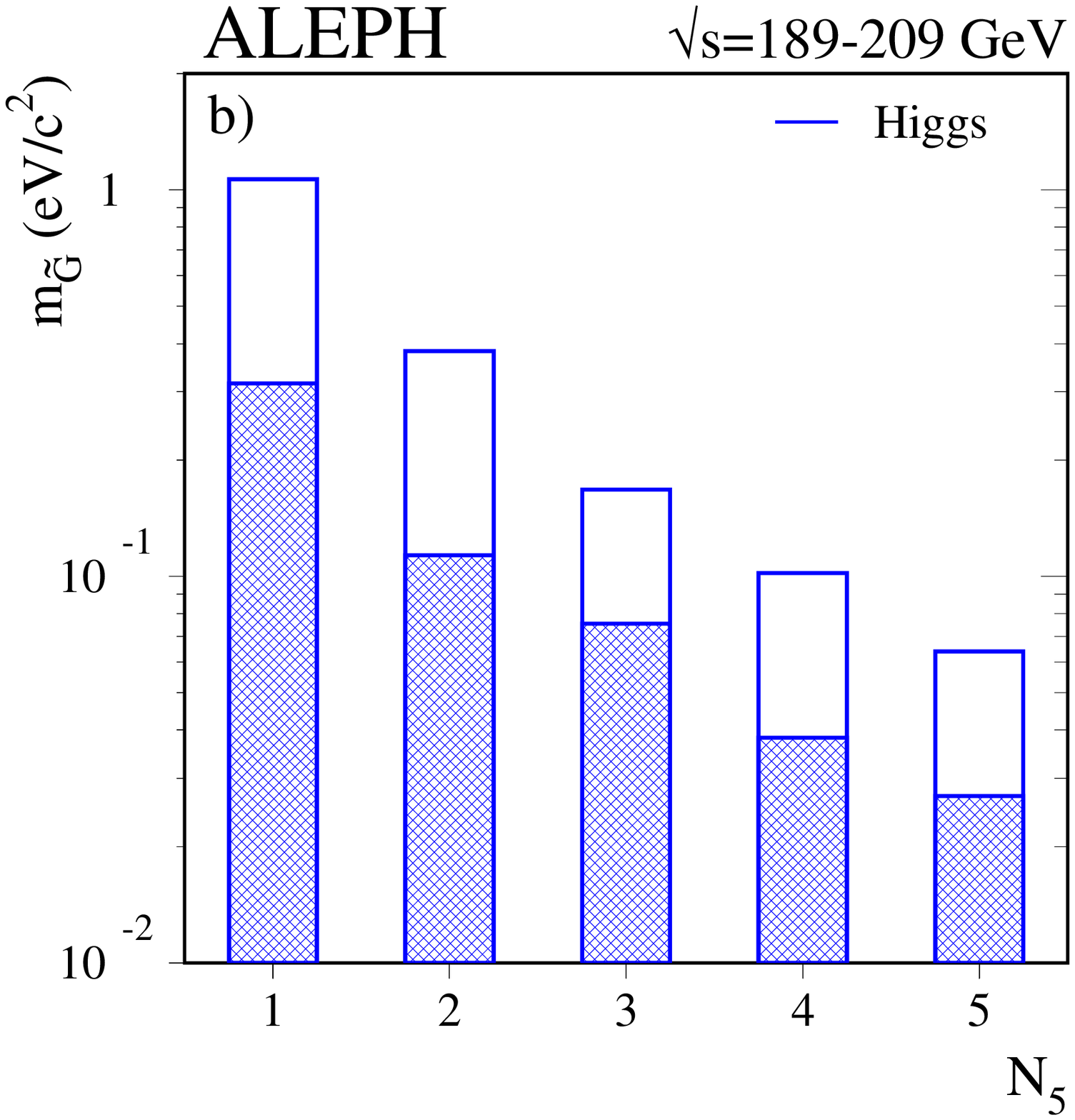,width=8cm}
\caption[]{\label{lambdan5}{\small Exclusions at 95$\%$ confidence level (a) for 
$\Lambda$ and (b) for $m_{\grav}$ as a function of $N_5$, derived from
the minimal GMSB scan (shaded). The unshaded bars represent the excluded
region when the neutral Higgs boson exclusion is applied with $\mtop = 175\gevcc$.}}
\end{center}
\end{figure}

The lower limit on $\Lambda$ independent of lifetime and $\tanb$ is shown in 
Fig.~\ref{lambdan5}a as a function of $N_5$. The absolute limit of
$\Lambda > 10\tevcc$ can be seen here for $N_5 = 5$.
When the Higgs boson search results are taken into account, the limit on $\Lambda$ increases to 16$\tevcc$, for $\mtop = 175\gevcc$. With a top mass
of 180$\gevcc$, the absolute lower limit on $\Lambda$ is set at 15$\tevcc$. 

The equation that relates the gravitino mass to the scale of SUSY
breaking in the messenger sector, $m_{\grav} = F /
\sqrt{3}M_{\mathrm{Planck}}$, can be exploited to put an indirect limit 
on the gravitino mass. 
The universal mass scale %($~\sim F/M_{\mathrm{mess}}$)
must obey $\Lambda \leq \sqrt{F}$ under the simple assumption of
positive messenger masses squared~\cite{theory}. The lower limit on
$\Lambda$ can then be converted into a constraint on $\sqrt{F}$ and
therefore provides an indirect limit on the gravitino mass.
The dependence of $m_{\grav}$ on $N_5$ is illustrated in
Fig.~\ref{lambdan5}b. The lower limit on $\Lambda$ of 10$\tevcc$ 
implies a lower limit on $m_{\grav}$ of 0.024$\evcc$. When the results of
Higgs searches are included these limits become 16$\tevcc$ and
0.061$\evcc$, respectively.

%%%%%%%%%%%%%%%%%%%%%%%%%%%%%%%%%%%%%%%%%%%%%%%%%%%%%%%%%%%%%%%%%%%%%%
% Conclusions
%%%%%%%%%%%%%%%%%%%%%%%%%%%%%%%%%%%%%%%%%%%%%%%%%%%%%%%%%%%%%%%%%%%%%%
\section{Conclusions}
No evidence for new physics has been found in the search for GMSB topologies
in the final ALEPH data sample collected at \roots\ up to 209\,GeV.
In order to test the impact of the searches reported here and in
Refs.~\cite{slepba,charg01,aleph_higgs}, a scan over a minimal set of GMSB
parameters has been performed. The resulting NLSP mass limits can be read in
Table~\ref{concl}.

The scan also provides a lower
limit of 16$\tevcc$ on the universal SUSY mass scale $\Lambda$ and an
indirect lower limit on the gravitino mass of 0.061$\evcc$. 
\begin{table}[tb]
\begin{center}
\caption{\label{concl} \small NLSP mass limits, as derived from the scan}
\vspace*{0.2cm}
\begin{tabular}{|c|c|l|} \hline
NLSP         &  mass limit (95$\%$ C.L.) & validity \\ \hline \hline
                         & 92 \gevcc     & short $\neu$ lifetime \\ 
\raisebox{1.9ex}[-1.5ex]{$\neu$} & 54 \gevcc     & any lifetime \\ \hline
$\stau_1$                & 77 \gevcc     & any lifetime \\ \hline
any                      & 77 \gevcc     & Higgs exclusion \\ \hline
\end{tabular}
\end{center}
\end{table}

\section*{Acknowledgements}
It is a pleasure to congratulate our colleagues from the accelerator
divisions for the successful operation of LEP at high energy.
We are indebted to the engineers and technicians in all our institutions 
for their contribution to the excellent performance of ALEPH. Those of us
from non-member states wish to thank CERN for its hospitality and support. 
%% APPENDICES:
%
% Needed for the table (also in App. B)
\newcommand{\ORtext}[1]{$\left.\rule{0mm}{#1}\right\}$or}
\newcommand{\OR}{\multirow{2}{0ex}{\ORtext{2.5ex}}}
\section*{Appendix A: Four leptons with negligible lifetime selection}
\label{eva_app} 

The experimental topology consists 
of two energetic (from the $\slep \rightarrow \ell \tilde{G}$ decay) and two 
soft (from the $\chi \rightarrow \ell \slep$ decay) leptons plus missing energy and
momentum. The charge of the two most energetic leptons is expected to be the same 
in 50$\%$ of the cases.

In the analysis the presence of at least two
energetic leptons is required, where, in the case of tau decays, a lepton can also be a jet with 
small multiplicity and invariant mass. The same muon, electron and tau identification as described in
\cite{oldcasc} is applied. 

All topologies have a common anti--$\gamma \gamma$ preselection,
based on the rejection of events with
low transverse momentum or with energy deposits at small polar angles, which indicate the
presence of a scattered electron. The anti--$\gamma \gamma$ cuts include:

\begin{itemize}
\item $p_{\perp}/\sqrt{s} > 0.075 ~\mbox{or}~
(p_{\perp}/\sqrt{s} > 0.05 ~\mbox{and}~ |\phi_{\mbox{\tiny miss}} -  
90^{\circ}| > 15^{\circ} ~\mbox{and}~   
|\phi_{\mbox{\tiny miss}} - 270^{\circ}| > 15^{\circ})$; 
\item $\theta_{\mbox{\tiny diff}} > 5^{\circ} ~\mbox{or}~ \theta_{\mbox{\tiny scatt}} > 15^{\circ}$;
\item $E_{\mbox{\tiny nh}}/E_{\mbox{\tiny tot}} < 0.45 ~\mbox{and}~ 
(E_{\mbox{\tiny nh}}/E_{\mbox{\tiny tot}} < 0.30 ~\mbox{or}~ 
p_{\perp \mbox{\tiny nnh}}/\sqrt{s} > 0.03)$;
\item $\cos \theta_{\mbox{\tiny miss}} < 0.95$
\end{itemize}

The cut variables are defined as follows:
$p_{\perp}$ is the transverse momentum of the event, 
$\phi_{\mbox{\tiny miss}}$ and $\theta_{\mbox{\tiny miss}}$ are the azimuthal and polar angles of the missing momentum, $\theta_{\mbox{\tiny diff}}$ and
$\theta_{\mbox{\tiny scatt}}$ are two angles associated with the 
$\gamma \gamma$ kinematic hypothesis described in Ref.~\cite{top},
$E_{\mbox{\tiny nh}}$ is the reconstructed neutral--hadron energy, 
$E_{\mbox{\tiny tot}}$ is the total reconstructed energy of the
event and $p_{\perp \mbox{\tiny nnh}}$ is the transverse momentum of the event 
evaluated without the neutral hadrons.

\subsection*{A.1 {\boldmath$\sel\sel$, $\smu\smu$,
$\sel\smu$} selections}

Events with 2, 3 or 4 charged tracks are considered in the case of 
topologies that do not involve tau leptons in the final state. 
At least three identified electrons (muons) are required in the $\sel\sel$
($\smu\smu$) selections. In the $\sel\smu$ selection the number of
identified leptons must again be at least three, but the two most energetic
leptons must have different flavours and no more than two leptons of the
same flavour are allowed.  These requirements reject most hadronic
background decays.
After the preselection, the cuts listed in Table~\ref{table1} are applied
to reject planar events and improve the $\gamma \gamma$ suppression. 
If the two most energetic leptons have different charges, further
selections are necessary to reduce the remaining WW and ZZ background, 
and in the case of $\sel\sel$, the Bhabha background.

\begin{table}[ht]
\center
\caption{\label{table1} \small Selection cuts applied to the $\sel\sel$,
$\smu\smu$ and $\sel\smu$ topologies.}
\vspace*{0.2cm}
\begin{tabular}{|c|c|c|c|} \hline
Variable & $\sel\sel$ & $\smu\smu$ & $\sel\smu$ \\
\hline
$E_{\mbox{\tiny 30}}/\sqrt{s}$ & $< 0.12$ & $< 0.12$ & $< 0.12$ \\
$\mbox{AcopT}$ & $< 175^{\circ}$ & $< 175^{\circ}$ & $< 175^{\circ}$ \\
$\mbox{Acol}$ & $< 176^{\circ}$ & - & - \\
$E_{\ell1}/\sqrt{s}$ & $< 0.37$ & - & - \\
\hline
\multicolumn{4}{|c|}{if $\mbox{Charge}(\ell1) \neq \mbox{Charge}(\ell2)$} \\
\hline
$E_{\ell1}/\sqrt{s}$ & $~~~~< 0.39~~~~$ & $< 0.40$ & $< 0.39$ \\
$E_{\ell2}/\sqrt{s}$ & $< 0.33$\OR & - & - \\
$\gamma_{\mathrm{iso}}$ & veto & - & - \\
$\mbox{Acol}$ & - & - & $< 172^{\circ}$ \\
$E_{\mbox{\tiny lt}}/\sqrt{s}$ & - & - & $> 0.12 ~\mbox{and}~ <0.37$ \\
\hline
\end{tabular}
\end{table}
\normalsize

The cut variables are defined as follows: $E_{\mbox{\tiny 30}}$ is the total energy
reconstructed in a $30^{\circ}$ cone around the beam line,
$\mbox{AcopT}$~\cite{stauLEP1} is the transverse acoplanarity of the two
most energetic leptons, $\ell1$ and $\ell2$.
$\mbox{Acol}$ is the angle between $\ell1$ and $\ell2$ in space and
$E_{\ell1}$ is the energy of the most energetic lepton in the event.
In the $\sel\sel$ case, an event is accepted if either the cut on the
energy of the second most energetic lepton, $E_{\ell2}$, or a veto on
isolated photons, $\gamma_{\mathrm{iso}}$, is satisfied. $E_{\mbox{\tiny
lt}}$ is the energy of the leading track in the event.

\subsection*{A.2 {\boldmath$\sel\stau$, $\smu\stau$,
$\stau\stau$} selections}

The selection procedures for the $\sel\stau$ and $\smu\stau$ topologies are
based on the acoplanar leptons analysis developed in Ref.~\cite{oldcasc}
where a detailed definition of the tau reconstruction algorithm is also given. The selection cuts for mixed topologies
are listed in Table~\ref{table2}.

\begin{table}[ht]
\center
\caption{\label{table2} \small Selection cuts applied to the $\sel\stau$ and $\smu\stau$ topologies.}
\vspace*{0.2cm}
\begin{tabular}{|c|c|c|} \hline
Variable & $\sel\stau$ & $\smu\stau$ \\
\hline
$N(\mbox{charged})$ & $=(4,5,6)$ & $=(4,5,6)$ \\
$N(\ell = \mu ~\mbox{or~e})$ & $N(\mbox{e}) \geq 1$ & $N(\mu) \geq 1$ \\
$N_{\tau}$ & $\geq 2$ & $\geq 2$ \\ 
$M_{\mbox{\tiny tot}}/\sqrt{s}$ & $> 0.06 ~\mbox{and}~ < 0.6$ & $> 0.06 ~\mbox{and}~ < 0.6$ \\
$\mbox{Acol}$ & $< 175^{\circ}$ & $< 175^{\circ}$ \\
$E_{\mbox{\tiny 12}}/\sqrt{s}$ & $< 0.02$ & $< 0.02$ \\
$\mbox{Thrust}$ & $< 0.96$ & $< 0.96$ \\
$M_{\mbox{\tiny miss}}/\sqrt{s}$ & $> 0.1  ~\mbox{and}~ < 0.8$ & $> 0.1 ~\mbox{and}~ < 0.8$ \\
$M(\mbox{Event} - \tau1)/\sqrt{s}$ & $< 0.54$ & $< 0.55$ \\
$E_{\tau1}/\sqrt{s}$ & $> 0.17 ~\mbox{and}~ < 0.38$ & $> 0.14 ~\mbox{and}~ < 0.36$ \\
$E_{\tau2}/\sqrt{s}$ & $> 0.01 ~\mbox{and}~ < 0.26$ & $> 0.01 ~\mbox{and}~ < 0.30$ \\
$M_{\tau1}/\sqrt{s}$ & $> 0.04 ~\mbox{and}~ < 0.25$ & $> 0.02 ~\mbox{and}~ < 0.25$ \\
$M_{\tau2}/\sqrt{s}$ & $< 0.17$ & $< 0.18$ \\
$E_{\ell1}/\sqrt{s}$ & $> 0.14 ~\mbox{and}~ < 0.36$ & $> 0.12 ~\mbox{and}~ < 0.36$ \\
$E_{\ell2}/\sqrt{s}$ & $< 0.24$ & $< 0.28$ \\
$N_{\mbox{\tiny nh}}(\tau1)$ & - & $< 2$ \\
\hline
\end{tabular}
\end{table}
\normalsize

The definition of the variables used is the following: $N(\mbox{charged})$
is the number of reconstructed charged tracks,
$N(\ell)$ is the number of identified leptons ($\ell$ = e, $\mu$),
$M_{\mbox{\tiny tot}}$ is the invariant mass of the event,
$\mbox{Acol}$ is the acollinearity of the two tau jets, $E_{\mbox{\tiny 12}}$ is the total energy
reconstructed in a $12^{\circ}$ cone around the beam line, $M_{\mbox{\tiny miss}}$ is the missing mass of
the event, $M(\mbox{Event} - \tau1)$ is the invariant mass of the event once the most energetic
tau jet, $\tau1$, has been removed, $E_{\tau1}$ and $E_{\tau2}$ are the energies of the
two tau jets, $M_{\tau1}$ and $M_{\tau2}$
are the corresponding masses. $N_{\mbox{\tiny nh}}(\tau1)$ is the number of
neutral hadrons reconstructed in the most energetic tau jet.

The cuts on $M_{\mbox{\tiny tot}}$ and $E_{\mbox{\tiny 12}}$ allow further 
$\gamma \gamma$ suppression while the cuts on $N(\mbox{charged})$,
$N(\ell)$ and $N_{\mbox{\tiny nh}}(\tau1)$ reject most of the
$q\bar{q}$ background and hadronic decays. The kinematic cuts on the
jet variables select events containing tau-like jets.

For the $\stau\stau$ topology
the events are clustered into four jets
using the Durham algorithm~\cite{durham} and each jet is checked for consistency with a
tau hypothesis. Events satisfying the cuts listed in Table~\ref{table3} are
retained.
Two alternative strategies are applied depending on wether the sum of the charges of the 
tracks in the two most--energetic tau jets is zero, $\mbox{Charge(jet1 + jet2)} = 0$, or
$\mbox{Charge(jet1 + jet2)} = 1 ~\mbox{or}~ 2$.

\begin{table}[ht]
\center
\caption{\label{table3} \small Selection cuts applied to the $\stau\stau$ topology.}
\vspace*{0.2cm}
\begin{tabular}{|c|c|c|} \hline
Variable & \multicolumn{2}{c|}{$\stau\stau$} \\
\hline
$-\ln y_{\mbox{\tiny 23}}$ & \multicolumn{2}{c|}{$> 2 ~\mbox{and}~ < 8.5$} \\
$-\ln y_{\mbox{\tiny 34}}$ & \multicolumn{2}{c|}{$< 9.5$} \\
$-\ln y_{\mbox{\tiny 45}}$ & \multicolumn{2}{c|}{$< 11$} \\
$N_{\mbox{\tiny good}}$ & \multicolumn{2}{c|}{$> 2 ~\mbox{and}~ < 9$} \\
$N_{\mbox{\tiny cj}}$ & \multicolumn{2}{c|}{$\geq 3$} \\
$N(\tau)$ & \multicolumn{2}{c|}{$\geq 3$} \\
$N_{\mbox{\tiny cnt}}$ & \multicolumn{2}{c|}{$\leq 2$} \\
$M_{\mbox{\tiny tot}}/\sqrt{s}$ & \multicolumn{2}{c|}{$> 0.06 ~\mbox{and}~ < 0.5$} \\
$E_{\mbox{\tiny 12}}/\sqrt{s}$ & \multicolumn{2}{c|}{$< 0.05$} \\
$E_{\mbox{\tiny 30}}/\sqrt{s}$ & \multicolumn{2}{c|}{$< 0.09$} \\
$\mbox{AcopT}$ & \multicolumn{2}{c|}{$< 178^{\circ}$} \\
$E_{\tau_{\mbox{\tiny max}}}/\sqrt{s}$ & \multicolumn{2}{c|}{$< 0.3$} \\
$M_{\mbox{\tiny miss}}/\sqrt{s}$ & \multicolumn{2}{c|}{$> 0.3 ~\mbox{and}~ < 0.9$} \\
\hline
$\mbox{Charge(jet1 + jet2)}$ & $= 0$ & $= 1,2$ \\
\hline
$\mbox{Thrust}$ & $< 0.96$ & $< 0.98$ \\
$\mbox{Acop}$ & $< 172^{\circ}$ & $< 171^{\circ}$ \\
$p_{\ell_{\mbox{\tiny max}}}/\sqrt{s}$ & $< 0.23$ & $< 0.26$ \\
\hline
\end{tabular} 
\end{table}
\normalsize

The cut variables are defined as follows: $y_{\mbox{\tiny 23}}$ 
($y_{\mbox{\tiny 34}}$, $y_{\mbox{\tiny 45}}$) is the $y$ cut for which 
the event goes from 2 to 3 (3 to 4, 4 to 5) jets. $N_{\mbox{\tiny good}}$ is 
the number of good tracks, $N_{\mbox{\tiny cj}}$ is the number of jets
containing at least one charged track, $N(\tau)$ is the number of identified 
tau jets, $N_{\mbox{\tiny cnt}}$ is the number of charged tracks not 
associated with any tau jet. $E_{\tau_{\mbox{\tiny max}}}$ is the energy of
the most energetic tau jet in the event. $p_{\ell_{\mbox{\tiny max}}}$ is 
the momentum of the most energetic identified lepton (e or $\mu$)
from the most energetic tau jet (including final state radiation).

The cuts have been optimised to select events with at least three tau-like 
jets not lying in the same plane. They reject most Standard Model
processes leaving an irreducible contribution arising from WW and $\tau
\tau (\gamma)$ events.

%
% Luke's defs:
\newlength{\lj}
\newcommand{\dzero}{\ensuremath{d_0}}
\newcommand{\zzero}{\ensuremath{|z_0|}}
\newcommand{\dl}{\ensuremath{d_{\slep}}}
\newcommand{\Nch}{\ensuremath{N_{\mathrm{ch}}}}
\newcommand{\Nchprim}{\ensuremath{\Nch'}}
\newcommand{\cch}{\ensuremath{c_{2ch}}}
\renewcommand{\Sb}{\ensuremath{S_{2\beta}'}}
\newcommand{\phiacoprim}{\ensuremath{\Phi_{\mathrm{aco}}'}}
\newcommand{\phiaco}{\ensuremath{\Phi_{\mathrm{aco}}}}
\newcommand{\Ntpc}{\ensuremath{N_{\mathrm{TPC}}}}
\newcommand{\Ntpcprim}{\ensuremath{\Ntpc'}}
\newcommand{\Etotprim}{\ensuremath{E_{\mathrm{tot}}'}}
\newcommand{\Etot}{\ensuremath{E_{\mathrm{tot}}}}
\newcommand{\mtotprim}{\ensuremath{m_{\mathrm{tot}}'}}
\newcommand{\mtot}{\ensuremath{m_{\mathrm{tot}}}}
\newcommand{\alphaprim}{\ensuremath{\alpha'}}
\newcommand{\alphax}{\ensuremath{\alpha}}
\newcommand{\Rl}{slepton hits}
\section*{Appendix B: Four leptons with lifetime selection} \label{luke_app}
In analogy with the negligible lifetime analysis, described in Appendix A,
six selections corresponding to the six different final states have been developed. 
\par
The main requirement for selection is that an event has at least one
{\it high-\dzero} track.
The main background to such an object is given by
secondary interactions of particles originating from SM processes with the 
detector material (e.g. nuclear interactions, photon conversions, etc). 
The rejection of this background is achieved through detailed analysis 
of the TPC, ITC and VDET hits associated to the  reconstructed
high-\dzero\ tracks. 
Three variables are used to select a high-\dzero\ track: the momentum,
the \dzero\ and $\chi^2_{\mathrm{IP}}$ (the $\chi^2$ of the track fit to the 
interaction point, normalised to the number of degrees of freedom).
The \dzero\ cut is not applied if the track \zzero\ is greater than 8\,cm. 
For each topology two sets of cuts have been applied to cope 
with the two cases of short slepton decay length ($\dl \sim 1$\,cm) 
and of long slepton decay length ($\dl \sim 1$\,m). 
These cuts are summarised in Table~\ref{trackcuts}. 
For the \sel\sel, \smu\smu\ and \sel\smu\ channels the electron and/or 
muon identification is also applied to the high-\dzero\ track.
\par
\begin{table}[ht]
\begin{center}
\caption{\label{trackcuts} \small The cuts on track momentum ($p$), 
\dzero\ and $\chi^2_{\mathrm{IP}}$ 
for each channel under each selection. A track must have parameters
greater than these values to be considered as a good high-\dzero\
track in the corresponding selection.}
\vspace*{0.2cm}
\begin{tabular}{|c|c|c|c|c|c|c|c|} \hline
Variable & Selection
&\sel\sel&\smu\smu&\stau\stau&\sel\smu&\sel\stau&\smu\stau \\ \hline
& low \dl  & 19.0 & 20.0 & 4.7 & 14.9 & 10.0 & 8.1 \\
\raisebox{1.5ex}[-1.5ex]{$p$ (GeV)}
& high \dl & 19.8 & 18.6 & 6.1 & 20.0 & 8.5 & 4.3 \\ \hline
& low \dl  & 0.17 & 0.50 & 0.20 & 0.20 & 0.24 & 0.50 \\
\raisebox{1.5ex}[-1.5ex]{\dzero\ (cm)}
& high \dl & 0.50 & 0.50 & 0.50 & 0.50 & 0.49 & 0.48 \\\hline
& low \dl  & 700 & 550 & 230 & 180 & 210 & 700 \\ 
\raisebox{1.5ex}[-1.5ex]{$\chi^2_{\mathrm{IP}}$} 
& high \dl & 680 & 690 & 700 & 700 & 590 & 610 \\ \hline
\end{tabular}
\end{center}
\end{table}
\par

In addition to the request of having at least one 
high-\dzero\ track, an event is also required to pass at least one of 
six selections based on global event variables. The principal cuts are 
listed in Table~\ref{globalcuts}.

\newcommand{\note}{\ensuremath{\dag}}
\newcommand{\snote}{\note\hspace*{1ex}} 
% dictates the space between the \note and the column separator to its right
\settowidth{\lj}{\snote}
\par
\begin{table}[ht]
\begin{center}
\caption{\label{globalcuts} \small Global event variable cuts.}
\vspace*{0.2cm}
\begin{tabular}{|c|c|c!{}@{\hspace*{-\lj}}c@{}|c|}
\hline
Variable        & \sel\sel, \smu\smu, \sel\smu & \sel\stau, \smu\stau & & \stau\stau\\
\hline                                                          
\Nch            & \gt2 and \lt20        & \gt2        &           &       --            \\
\Nchprim        & \gt7                  & \lt9        &           &       --            \\
\Ntpc           &  --                   & \lt12       &           & \gt2 and \lt14      \\ 
\Ntpcprim       &  --                   & \lt10       &           & \lt11               \\ 
\Etotprim       &  --                   & \gt6\ GeV   & \snote    & \gt6\ GeV\ and \lt0.75$\sqrt{s}$ \\
\mtotprim       &  --                   & \gt7.7\ GeV & \snote    & \gt7.7\ GeV          \\ 
\alphaprim      &  --                   & \lt178\degs & \snote    & \lt178\degs         \\ \hline
\cch            & \gt-0.999 and \lt0.99 & \gt-0.999 and \lt0.99 & & \gt-0.999 and \lt0.99 \\      
\Sb             & \gt0.1 \OR            & \gt0.1 \OR  &           &  \gt0.2\OR          \\      
\phiacoprim     & \lt174\degs           & \lt174\degs &           &\lt174\degs          \\\hline
\end{tabular}   
\end{center}
\end{table}   

Most analyses use only tracks with a \dzero\ less than 2\,cm\ and a
\zzero\ less than 10\,cm, and ignore all others. 
Some of the variables, indicated as primed, are
calculated using only tracks with these \dzero\ and \zzero\
conditions, while unprimed variables are calculated using all
tracks.
The definitions of the variables are as follows: \Nch\ is the
number of charged tracks, \Ntpc\ is the number of charged tracks with
at least one TPC hit, \Etot\ is the total energy of the event, \mtot\ is the
invariant mass of the event, \alphax\ is the acollinearity of the
event, \cch\ is cosine of the angle between the two highest momentum
tracks, \Sb\ is $\sqrt{1-0.5(\beta_1^2+\beta_2^2)}$ (where $\beta_1$
and $\beta_2$ are the boosts of the two event hemispheres) and
\phiaco\ is the acollinearity of the event.
\par
The cuts marked with a \note\ are not applied if a parent slepton
track has been identified (i.e.~a mother-daughter relationship has
been established between one of the high-\dzero\ tracks and a
track from the primary interaction point). The cuts on
\cch, \Sb\ and \phiacoprim\ are only applied in the case that there is
just one good high-\dzero\ track, and that no track or hit has been
tagged as belonging to its parent slepton. Only one of the cuts
grouped with a brace need to be passed. In addition, cosmic-ray events
are suppressed by requiring that the event be within 100\,ns of the
bunch crossing. 

%
% BIBLIOGRAPHY
%

\end{document}